\documentclass[12pt,letterpaper]{article}
\usepackage{amssymb,amsmath,amsthm,amsfonts,mathtools,mathrsfs,bm}
\usepackage[top=1in,bottom=1in,left=1in,right=1in]{geometry}
\usepackage[round]{natbib}
\usepackage{setspace}
\usepackage{xcolor}

\usepackage{graphicx}
\usepackage{booktabs,threeparttable,tabularx,array,multirow,makecell}
\usepackage{enumitem}
\usepackage{subcaption}
\newcommand{\eventnum}[1]{#1}
\usepackage[hyphens]{url}
\usepackage{xr-hyper}
\usepackage{hyperref}
\hypersetup{colorlinks=true,linkcolor=blue,citecolor=blue,urlcolor=blue,
            breaklinks=true,hypertexnames=false}
\usepackage{etoolbox}
\patchcmd{\abstract}{\small}{\normalsize}{}{}
\renewcommand{\footnotesize}{\normalsize\onehalfspacing}
\usepackage{float}
\usepackage{placeins}
\usepackage{tikz}
\usetikzlibrary{positioning,arrows.meta}
\makeatletter\let\sv@bibcite\bibcite\let\bibcite\@gobbletwo\makeatother
\makeatletter\let\bibcite\sv@bibcite\makeatother

\title{Equilibrium Transition and Cartel Formation:\\
       A Structural Analysis of Chile's Pharmacy Cartel}
\author{Yu (Jasmine) Hao\thanks{I thank the Chilean Competition Authority (FNE) for data access; the editor and anonymous referees for comments; and my advisor, Hiro Kasahara, my dissertation committee, and Victor Aguirregabiria for discussions. I also thank Jean-Fran\c{c}ois Houde for encouraging this study of the advertisement ban. I acknowledge support from the 2020 Gambling Award Fellowship. All errors are my own.}
\\[2pt]
\normalsize Faculty of Business and Economics\\ University of Hong Kong \\
\normalsize \texttt{haoyu@hku.hk}}
\date{\today}

\begin{document}
{\singlespace
\maketitle
}
\vspace{-2\baselineskip}

\begin{abstract}\noindent
This paper studies how Chile's three largest pharmacy chains moved from the price war to collusion, using court-record daily prices and a structural model. After a court-ordered advertisement ban ended the comparison campaign, the estimated gain from loss-leader pricing fell sharply, weakening the incentive to continue the price war. The chains then used an upstream supplier as intermediary to \textbf{verify} a collusive price-leadership \textbf{procedure}. Once verified, they applied it to restore margins on former loss leaders. They then raised prices further to extract rents. These increases spread more slowly than margin restoration. The Adaptive Confidence specification allows a subjective belief about follower participation to update. It more closely matches the cumulative spread of rent extraction than the Unit Confidence specification. Unit Confidence imposes the rational-belief benchmark by fixing the leader's weight at one.


\end{abstract}

\section{Introduction}\label{sec:intro}

How do firms start colluding? Although collusion has long been a central topic
in industrial organization, most studies focus on its implementation and
maintenance, with less attention to cartel formation. \citet{Harrington2018}
distinguishes the problem of coordinating on a collusive agreement from that
of ensuring compliance with it. Firms must establish a common understanding
of what each will do and believe that their rivals will act accordingly.
\citet{byrne2019learning} provide one of the few empirical studies of this
process, documenting how firms gradually establish a focal pricing cycle
through price leadership and experimentation. I use a structural
model to distinguish verification of a price-increase procedure from
willingness to participate in subsequent increases. Once firms have verified
that the procedure works, they still face uncertainty about their rivals'
willingness to participate. My analysis is also related to
\citet{igami2022measuring}, who use a structural model to measure firms'
incentives to sustain collusion and explain cartel breakdown. Here, the focus
is on the build-up of collusion.

I study Chile's three largest pharmacy chains: Cruz Verde, Farmacias Ahumada
(FASA), and Salcobrand. Together, they accounted for approximately 92 percent
of national retail pharmacy sales in 2007.\footnote{The cartel case is
\emph{Fiscal\'ia Nacional Econ\'omica v. Farmacias Ahumada S.A. and Others},
Rol C-184-08, before Chile's Competition Tribunal (TDLC). Judgment
No.~119/2012 sanctioned Cruz Verde and Salcobrand; FASA had previously
settled with Chile's competition authority, the Fiscal\'ia Nacional Econ\'omica (FNE) \citep{tdlc2012sentencia}. The FNE helped me access
the public case materials through Chile's transparency procedure.}
I use the case records together with daily prices and quantities,
wholesale costs, and manufacturers' suggested retail prices to study this
transition.
In late 2006, the chains began undercutting one another. They used loss-leader
drugs to attract customers who would buy the discounted medicine and other
medicines or products during the same pharmacy visit
\citep[consids.~42--43]{tdlc2012sentencia}. In August 2007, Cruz Verde
launched a campaign comparing its prices with FASA's. Its rivals responded
with further price cuts \citep{fne2008requerimiento}.

A court-ordered advertisement ban ended the comparison
campaign on 6 November 2007. The sustained price war ended at the same time.\footnote{Santiago's 17th Civil Court
ordered the campaign to stop in \emph{Farmacias Ahumada S.A. v. Farmacias
Cruz Verde S.A.}, case Rol C-23.423-2007
\citep{fne2008requerimiento,Juzgado17Civil2007Precautoria}.}
On 9 December 2007, with help from upstream suppliers, the chains first
successfully used a ``one-two-three'' price-increase procedure.
One chain raised its price first, and the other two followed in turn.
The suppliers passed the proposed prices and dates between
the chains \citep{fne2008requerimiento,tdlc2012sentencia}.
The chains quickly used the same procedure for other medicines,
raising prices toward the manufacturers' suggested retail levels and restoring
positive margins. Prices rose from below wholesale cost to approximately
22--25 percent above it.
\textbf{Further increases, which I call rent extraction,} raised prices to
approximately 43--48 percent above wholesale cost, but spread more slowly.
The competition authority began requesting information in late March 2008.

Before the first successful coordinated increase, the chains repeatedly tried and failed to raise prices together.
Between 6 November and 8 December 2007, the price records show 42 failed medicine-level attempts, 39 started by Salcobrand.
Salcobrand also asked pharmaceutical laboratories to help the chains coordinate increases
\citep[consids.~95, 97]{tdlc2012sentencia}.
Once the chains succeeded on 9 December, they quickly repeated the procedure across other medicines to restore positive margins.
In a 19 December email, Salcobrand described the completed sequence as a ``procedure'' and said it expected to repeat it with more products and laboratories
\citep[consid.~95]{tdlc2012sentencia}.

The chains then used the same procedure to raise prices further, moving beyond margin restoration to rent extraction.
These increases spread more slowly, even though the chains had already used the procedure successfully.
Over matched 65-day windows, the rate of first increases per eligible medicine-week was 1.83 times as high for margin restoration as for rent extraction (Section~\ref{ssec:desc_second_round}).
Why did further increases spread more slowly once the procedure was established?
A chain raising first still risked losing customers if its rivals did not follow.
I distinguish learning that the procedure works from learning whether rivals are willing to participate in further increases.

I model the advertisement ban as changing two primitives. The two primitives are demand and the set of pricing procedures available to firms. I estimate a logit demand model that allows a chain to attract extra demand by offering a low price. The estimated response to ordinary price differences changes little after the advertisement ban, but the extra gain from offering a low price falls sharply. The low-price group consists of chains pricing a medicine sufficiently below the highest-priced chain. More than one chain can belong to this group. The outside good is not purchasing the medicine from any of the three chains. Holding current prices and recent price changes constant, membership in the low-price group raises purchase odds relative to the outside good by 23.0 percent before the ban and 6.4 percent afterward. Demand also responds to recent price changes. A chain loses more customers after raising its price if its rivals have not followed, creating a cost of moving first. Recent cuts have no clear additional effect after controlling for current prices and low-price-group membership. I use these demand estimates to calculate the gains and losses from changing prices, including complementary basket profit from other purchases associated with the additional medicine demand.

I develop a dynamic pricing model with three blocks. The blocks are ordinary price competition, the advertisement campaign, and collusive price leadership. The advertisement campaign allows larger cuts and changes how often medicines are reviewed for a price adjustment. After the ban, laboratories provide opportunities to coordinate price increases. Before verification, Salcobrand knows the procedure while the other chains evaluate increases under ordinary price competition. After verification, one chain raises first and the other two decide in turn whether to follow. The leader's risk is that rivals decline and customers leave. Knowing the procedure does not tell the leader whether its rivals will join a further increase.

I compare two models of leader confidence in follower participation. Adaptive Confidence updates the weight the leader places on the expected profit from sequential follower choices relative to raising alone after each rent-extraction attempt. Unit Confidence fixes that weight at one. Followers choose whether to join in both models. I estimate four economic parameters in each model and two additional learning parameters in Adaptive Confidence by simulated method of moments, matching nine groups of moments to the data \citep{goettler2011amd,yang2020vertical}. By the end of the estimation period, Adaptive Confidence predicts that about 68 medicines reach the rent-extraction level, compared with 70 in the data and about 106 under Unit Confidence.

This article is related to several strands of literature. First, it contributes to the literature on cartel formation. Much of the collusion literature studies whether firms can sustain a profitable agreement \citep{Bain1959,GreenPorter1984}. Forming a cartel also requires firms to coordinate on what to do \citep{green2014tacit,Harrington2018}. \citet{byrne2019learning,byrne2026negotiating} document how firms move toward collusion through price leadership, experimentation, and bargaining through prices. Related studies examine how communication, pricing rules, and intermediaries help firms coordinate \citep{GenesoveMullin1999,genesove2001rules,Harrington2017PartialMutual,GerlachNguyen2021,miller2021oligopolistic}. \citet{chaves2025inner} study how fuel distributors helped coordinate price increases among gas stations, while \citet{AleChiletAtal2020} show how a trade association raised physicians' prices by coordinating their exit from insurer networks. Studies of the Chilean pharmacy cartel document the initial use of safer, more differentiated products, consistent with mistrust \citep{AleChilet2016}, and examine the costs and allocation of price leadership \citep{AleChilet2018}. I build on this work by separating verification of a price-increase procedure from uncertainty about rivals' willingness to participate. The same procedure spread quickly during margin restoration but more slowly during rent extraction. The structural approach is also related to \citet{igami2022measuring}, who study incentives to sustain collusion under a known quota agreement. My focus is on the build-up of collusion, when firms first verify the procedure and then consider further increases while participation remains uncertain.

Second, this paper relates to the literature on loss-leader pricing. Loss-leader pricing uses below-cost products to attract customers who also buy other goods \citep{lal1994retail,degraba2003volume,chen2012loss}. Profits from these cross-category purchases give the chains a reason to keep cutting medicine prices despite negative margins on the discounted medicines \citep{thomassen2017multicategory}. I connect this mechanism to the literature on advertising restrictions \citep{benham1972advertising,cady1976estimate,MilyoWaldfogel1999}. I use the demand estimates to assess how the ban weakened the incentive to continue the price war. The demand estimates show that the response to ordinary price differences changed little after the ban, while the extra demand gained from offering a low price fell sharply. The ban therefore reduced the gains from using low-priced medicines to attract customers, weakening the incentive to undercut.

Third, this paper relates to the literature on firm learning. The firm-learning literature studies learning about market conditions and competitors' behavior \citep{aguirregabiria2020firms}. \citet{doraszelski2018just} study how firms learn in a new electricity market using models of fictitious play and adaptive learning. Related work examines how retailers learn to set prices after liquor privatization \citep{huang2022learning} and how learning about demand affects investment in container shipping \citep{jeon2022learning}. \citet{aguirregabiria2020identification} allow firms' beliefs about rivals' actions to differ from actual behavior in a dynamic game of retail-chain location. I distinguish learning how a price-increase procedure works from learning whether rivals will participate. In my model, verification makes the procedure common knowledge, but uncertainty about participation remains. Under Adaptive Confidence, the leader uses a Beta-shaped outcome-updating rule for its subjective weight after each rent-extraction attempt. Complete participation raises the weight, while incomplete participation lowers it.\footnote{The reduced-form rule treats each selected rent-extraction attempt as a binary outcome: $y=1$ for complete participation and $y=0$ for incomplete participation. It starts from the pseudo-counts $\nu^r m_0^r$ and $\nu^r(1-m_0^r)$, so after $S_t^r$ complete and $F_t^r$ incomplete outcomes it produces the weight in Equation~\eqref{eq:mech_willingness_belief}. The resulting weight governs the leader's valuation; it is not the actual probability that both rivals follow.} Firms can therefore understand the same procedure yet remain uncertain about whether others will participate in further increases.

Section~2 presents the institutional setting and data. Section~3 documents the transition from the price war to collusion. Section~4 estimates demand and the return to undercutting. Section~5 presents and estimates the dynamic pricing model, separating procedure verification from beliefs about rival participation. Section~6 concludes.

\section{Institutional Setting and Data}\label{sec:background_data}

\subsection{Market overview}\label{ssec:background_industry}

The Chilean pharmacy industry was dominated by three chains. In 2007, Cruz
Verde, the largest chain,\footnote{Cruz~Verde was vertically integrated with the distributor Socofar
\citep{fne2008requerimiento}; see Online Appendix~\ref{app:direct_evidence},
``(b)~Three chains setting one national price schedule.''} accounted for $40.6\%$
of nationwide pharmacy sales, followed by FASA with
$27.7\%$ and Salcobrand with $23.8\%$.\footnote{The shares are the FNE's,
citing IMS Health Chile.  See Online Appendix~\ref{app:direct_evidence},
``(b)~Three chains setting one national price schedule.''} A contemporaneous
industry report counted $494$ Cruz~Verde locations, including $153$ franchises,
$359$ FASA locations, and $314$ Salcobrand locations
\citep[pp.~21, 41, and 55]{DuranKremerman2007}. The FNE reported
that most of Chile's roughly $600$ independent pharmacies did not carry branded
prescription medicines and therefore provided little competition to the three
chains for those products \citep[para.~82]{fne2008requerimiento}. Each chain centrally
set a national price schedule, so major price increases and cuts were implemented
chain-wide rather than store by store. All three chains sold many
products besides medicines. A low medicine price could therefore attract shoppers
who also bought other items.

According to the FNE, consumers could fill a prescription at any chain but
needed medical supervision to switch medicines. The physician therefore largely determined the medicine, while the
consumer chose the pharmacy. The FNE describes these consumers as ``captive through
the prescription''
(\emph{cautivos a trav\'es de la receta m\'edica})
\citep[paras.~42--44 and 74]{fne2008requerimiento}.\footnote{Online
Appendix~\ref{app:direct_evidence}, ``(c)~The FNE's account of prescription constraints.''} A $2006$ report by Chile's consumer-protection agency (SERNAC)
says that proximity was the main factor in
pharmacy choice, followed by price, loyalty to a chain, and comparing prices before
buying \citep{SERNAC2006PharmacyStudy}. Whether covered by public or private
health insurance, households still paid directly for retail medicines
\citep{CastilloLabordeVillalobos2013}.
Medicines were the largest component of household
out-of-pocket health spending \citep{Debrott2007}. I therefore model consumers' choice among chains for a given medicine.

Chile had separate public and private procurement channels. Chile's public medicine
procurement agency aggregated the requirements of public hospitals and primary-care
providers and procured medicines for the public health network \citep{CENABAST2007}.
Private pharmacy chains instead
purchased through commercial relationships with pharmaceutical laboratories and
distributors. A 2001 competition ruling required manufacturers to publish medicine
price lists and the criteria for volume and early-payment discounts. Because all three
large chains qualified for the maximum volume discounts, the FNE characterized their
acquisition (wholesale) costs as largely uniform and transparent \citep{fne2008requerimiento}.
The same manufacturer lists included a common suggested retail price for each medicine,
known as the PVPS (\emph{precio de venta a p\'ublico sugerido}). Each chain centrally
set its own national retail price schedule. According to the FNE, the PVPS generally gave the chains a gross
margin of $20$--$25\%$ and provided a common benchmark for profitable retail
pricing.\footnote{Online
Appendix~\ref{app:direct_evidence}, ``(d)~The PVPS was a common,
manufacturer-supplied benchmark,'' reproduces the FNE's description of the PVPS
and the associated margin. The tribunal left this margin
characterization unresolved.}

\subsection{From price war to coordinated increases}\label{ssec:background_phases}

The PVPS was only a benchmark. Each chain could price below it.  The price record
passes through four episodes.  First, through September~$2006$, the chains offered
regular Monday and Thursday discounts that normally reversed within three days.
These were temporary promotions rather than a sustained reduction in the price
level.

Second, the temporary promotions became persistent cuts.  Salcobrand stopped
reversing its discounts on $22$~October~$2006$, Cruz~Verde did so on
$23$~November, and FASA lowered its prices more gradually.  From then through
mid-$2007$, the chains repeatedly undercut one another and average prices
declined.\footnote{Online Appendix~\ref{app:direct_evidence},
``(e)~Temporary discounts became persistent cuts at the end of $2006$,''
documents the break in the daily price record and reports the validation figure.}
The chains used high-turnover branded medicines as loss leaders, trading losses on the discounted medicines for profits on other purchases during the same pharmacy visit \citep[consids.~42--43]{tdlc2012sentencia}.

Third, Cruz~Verde formalized the price war in August~$2007$ with a campaign that
compared its prices on $685$ high-turnover branded medicines with FASA's prices.%
\footnote{Online Appendix~\ref{app:direct_evidence}, ``(f)~Comparative
advertising intensified the price war in August~$2007$,'' gives the
complaint's description, the press record, and the uncontested finding that the
price war compressed margins and produced below-cost sales.}
By publicizing the price difference, comparative advertising allowed the
lowest-priced chain to attract shoppers from its rivals.
FASA and Salcobrand responded by cutting prices on the advertised medicines to
defend their market shares, pushing many prices below wholesale cost.  FASA asked
the Council for Advertising Self-Regulation and Ethics (\emph{Consejo de
Autorregulaci\'on y \'Etica Publicitaria}, CONAR) to stop the campaign.  CONAR
ruled against Cruz~Verde on $7$~September and upheld that ruling on $5$~October.
Both decisions were nonbinding.

Fourth, FASA obtained a binding advertisement ban from Santiago's 17th Civil
Court on $6$~November.\footnote{Online
Appendix~\ref{app:direct_evidence}, ``(g)~The advertisement ban became binding on
$6$~November~$2007$,'' documents the date and FASA's account of why further cuts stopped making sense \citep{Juzgado17Civil2007Precautoria}. Figure~\ref{fig:undercut_quantity} plots the relative-quantity response around an undercut.}
The binding ban ended the comparison campaign, removing the channel through which a chain could publicize its low prices and attract shoppers. FASA's chief executive consequently described further price cuts as no longer making sense, and the sustained price war ended at the same time. Individual price cuts continued. The chains then used a separate laboratory-mediated procedure to
sequence price increases, beginning in December~$2007$. The FNE's requests for
information reached the chains at the end of March~$2008$, after which new
coordinated increases slowed.
A few more coordinated increases occurred in April, followed by a pause.
Coordinated increases resumed in July, while prices that had already been
raised largely stayed at their higher levels.

\begin{figure}[htbp]
\centering
\caption{From price war to coordinated price increases}
\label{fig:institutional_timeline}
\begin{subfigure}{\linewidth}
\centering
\includegraphics[width=\linewidth]{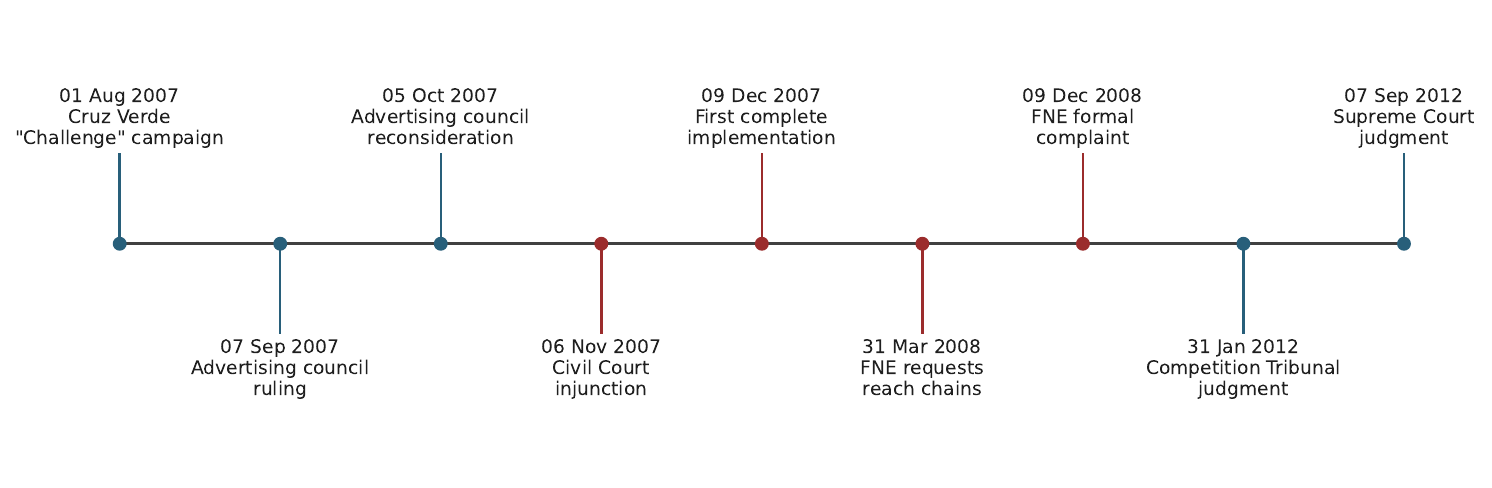}
\caption{Institutional timeline}
\label{fig:institutional_timeline_a}
\end{subfigure}
\par\medskip
\begin{subfigure}{0.92\linewidth}
\centering
\includegraphics[width=\linewidth]{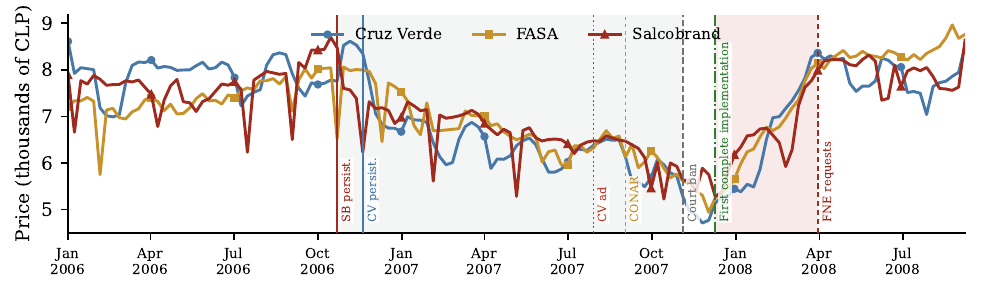}
\caption{Weekly average prices by chain}
\label{fig:institutional_timeline_b}
\end{subfigure}
\par\smallskip\flushleft
{\footnotesize \textbf{Note}: Panel~(b): quantity-weighted weekly prices across 222 medicines, in thousand
CLP. Gray shading denotes the price war; red denotes the initial cartel wave
(9~December~2007--31~March~2008). CV: Cruz Verde; SB: Salcobrand;
``persist.'': persistent cuts. Vertical lines mark the events discussed in the text.}
\end{figure}

Figure~\ref{fig:institutional_timeline} places the institutional dates alongside
the weekly average-price series.  Panel~(\subref{fig:institutional_timeline_a})
reports the legal and case chronology.  Panel~(\subref{fig:institutional_timeline_b})
shows the stable promotional period, the decline during persistent undercutting,
and the price recovery during the cartel wave (9~December~2007--31~March~2008).

\FloatBarrier

\subsection{Laboratory-mediated coordination}\label{ssec:coordination_mechanism}

The evidence below traces how the chains established a common way to sequence and monitor price increases, then expanded it after the first successful batches. At first, they lacked this common procedure. Laboratory mediation resolved that coordination problem without removing the leader's uncertainty about whether rivals would participate in further increases.
The price war left many medicines below wholesale cost.
Restoring margins required one chain to raise first without knowing whether
both rivals would follow. In its defence, Salcobrand stated that its new
management sought to restore margins even at the risk of losing market share.
Salcobrand offered to raise first, leaving FASA and Cruz~Verde three or four
days to detect and match each increase.\footnote{Online
Appendix~\ref{app:direct_evidence}, ``(h)~Salcobrand's new management sought to
restore margins,'' documents Salcobrand's stated margin strategy.
``(k)~The one-two-three procedure: Day One, Day Two, Day Three,'' reproduces the
internal email describing its offer to move first.}

Pharmaceutical laboratories became intermediaries in the
one-two-three procedure. Their managers communicated proposed prices and dates,
informed the other chains after the first chain moved, and checked whether all
three had followed. The chains monitored compliance through
price checks by their own or contracted staff and reports from the
laboratories.\footnote{Online Appendix~\ref{app:direct_evidence},
``(l)~Monitoring: prior knowledge and quotation intensity,'' documents these
monitoring channels; see also \citealp[para.~96, p.~35]{fne2008requerimiento}.}
When either rival did not follow, the initiator typically
returned to its previous price. A Bayer batch illustrates the sequence.
Salcobrand raised the full list on Day One, and FASA and Cruz~Verde each raised
half on Day Two and the remainder on Day Three
\citep[consid.~75]{tdlc2012sentencia}.\footnote{Online
Appendix~\ref{app:direct_evidence}, ``(j)~What the laboratories transmitted,''
reproduces the complaint's description, one laboratory email, and one
laboratory-manager statement.}

The case record contains conflicting accounts of who first
proposed the procedure. FASA and Salcobrand attributed the proposal to the
laboratories, whereas laboratory managers said that the chains asked them to
contact rivals and coordinate dates. The TDLC read the record as showing that
the push for coordination came from the pharmacy side. A Salcobrand email dated
$19$~December reported successful early batches for five products involving
four laboratories and expected the procedure to be repeated with more products
and laboratories.\footnote{Online Appendix~\ref{app:direct_evidence},
``(i)~Who proposed the coordination,'' presents the conflicting accounts and
the tribunal's reasoning. The Salcobrand email of $19$~December~$2007$ is
reproduced in ``(n)~The agreement expanded after successful early batches'';
see also \citealp[consids.~95 and 97]{tdlc2012sentencia}. Online Appendix
Section~\ref{app:lab_batches}, ``Laboratory batches,'' documents
the timing using the three chains' daily transaction-price paths.}

The three chains' daily transaction-price paths record the first
complete three-chain implementation on $9$~December~$2007$. I use this
observed implementation to date the model's transition to a commonly understood procedure. Under the weekly information transition
in Section~\ref{ssec:mech_stage}, informed play begins the following week. The FNE's requests for information reached the chains
at the end of March~$2008$. The formal notice to Salcobrand was Oficio
No.~$419$. The FNE filed its formal
complaint on $9$~December~$2008$, and the TDLC later sanctioned Cruz~Verde and Salcobrand
for coordinating prices on at least \eventnum{$206$} medicines
\citep[consids.~191, 205--206]{tdlc2012sentencia}.\footnote{Online
Appendix~\ref{app:direct_evidence}, ``(o)~Scope, chronology, and outcome,'' and
``(p)~Communication after the investigation began.'' The
judgment dates Oficio No.~$419$, addressed to Salcobrand, to
$3$~April~$2008$; testimony places the FNE's requests at the end of March. The
judgment's ``at least $206$ medicines'' is a legal minimum for
December~$2007$--March~$2008$, not the event count defined below.}

The chains argued that the price increases reflected independent
responses to a price leader rather than collusion. The tribunal acknowledged
that the price patterns alone could support either explanation. It found
collusion after considering the emails and testimony alongside the price
evidence.\footnote{Online Appendix~\ref{app:direct_evidence},
``(q)~The alternative explanations on the record,'' presents the competing
accounts and the tribunal's assessment; see also
\citealp[consids.~159, 165--166, 192]{tdlc2012sentencia}.}

\subsection{Data and event construction}\label{ssec:background_data}

\noindent\textbf{(i)~Prices and quantities.} I use a processed
daily panel covering 222 medicines at Cruz~Verde, FASA, and
Salcobrand from 2006 through 2008. Each chain--drug--day observation contains a
revenue-weighted average transaction price and units sold. Separate listed-price
series cover the same panel. I use transaction prices to date events and to
measure realized prices, markups, price tiers, and demand. Salcobrand's
wholesale costs, available from November~$2007$ onward, come from evidence
submitted to the TDLC.\footnote{Online Appendix Section~\ref{app:data}, ``Data construction,'' describes the data and sample selection.}

\noindent\textbf{(ii)~Event panel.} I classify persistent
drug-level price plateaus into three tiers. Tier~0 is the
below-cost tier, with quantity-and-cost-weighted cost-relative markups of
$9$--$12\%$ below wholesale cost across chains. Tier~1 is the
margin-restoration tier, the first common price that restores a positive retail
margin, with markups of $22$--$25\%$ above wholesale cost. Tier~2 is the
rent-extraction tier, the next common price above Tier~1, with markups of
{$43$--$48\%$} above wholesale cost. These markups use wholesale cost as the
denominator, $(p-c)/c$, rather than the retail-price margin defined above.
Table~\ref{tab:desc_tier_markups} reports the chain-specific values.
For each medicine I compute one benchmark price per tier,
common across the three chains, by averaging their weekly transaction prices
over the tier's window. Tier~0 uses the price-war period. Tiers~1 and~2 use
stable periods beginning two weeks after the corresponding first successful
increase. Where a benchmark price is missing, I use the median
Tier~1-to-Tier~0 price ratio across medicines for Tier~1. For Tier~2, I use a markup
regression estimated on medicines with an observed stable Tier~2 window for
Tier~2.\footnote{Online Appendix Section~\ref{app:event_coding}, ``Construction of price tiers,'' explains how I calculate the benchmark tier prices.}
I construct the event panel from the three chains' daily
transaction-price paths. The detection procedure flags sustained changes
relative to a chain's recent price and its rivals' prices. Coordinated-increase
candidates are then checked for their date, initiator, rival response, and
persistence. The panel records successful and unsuccessful increases and price
cuts before and after coordination begins.

An increase event is one medicine-level attempt, including the initiating
chain's price change and the rivals' responses. It is successful when all three chains reach their chain-specific prices within
the same higher tier. The attempt is unsuccessful if the third chain does not follow. For each medicine, I record the first time its
price crosses each adjacent tier boundary. A direct Tier~$0\to2$ increase is
therefore one event, but the observed boundary counts record it as the medicine's first
crossing of both the Tier~$0\to1$ and Tier~$1\to2$ boundaries.

\noindent\textbf{(iii)~Documentary record.} I use the FNE
complaint, the TDLC decision, the CONAR ruling, and the civil-court injunction
to reconstruct the institutional timeline and communication among the chains.\footnote{Online Appendix~\ref{app:direct_evidence}, ``(a)~Sources and evidentiary conventions,'' identifies the document supporting each institutional fact.}

\begingroup
\section{Descriptive Evidence}\label{sec:descriptive}

This section documents four patterns used in the demand system
and dynamic game. Temporary discounts became persistent. The quantity response
to an undercut changed after the advertisement ban. Increase attempts became
organized by laboratory and date. The movement from the below-cost price to
the first common price with a positive retail margin (Tier~$0\to1$) was faster
than the subsequent movement from that positive-margin price to the next higher
common price (Tier~$1\to2$). The model takes the start of the price war as given.

\subsection{From temporary discounts to persistent low prices}\label{ssec:desc_persistence}

For most of~$2006$, discounts at Cruz~Verde and Salcobrand
reversed within days. Among January--September markdowns of at least $15\%$,
$96.4\%$ at Salcobrand and $95.2\%$ at Cruz~Verde returned to at least $95\%$
of the pre-markdown price within three days. This pattern changed late in the
year. Salcobrand cut 129 medicines by a median of $21\%$ on 22~October, and
only one returned to that threshold the next day. Cruz~Verde cut 164 medicines
by a median of $24\%$ on 23~November, and none returned the next day. By late
November, the share of medicines with weekly median prices at least $10\%$ below their January--September median of weekly median prices exceeded one-half at both chains in each of four consecutive weeks (Online
Appendix Figure~\ref{fig:discount_validation}, Panel~(c)).\footnote{The
markdown counts require consecutive-day cuts of at least $15\%$. Prices are
observed on both days for 189 Salcobrand medicines and 209 Cruz~Verde
medicines. Contemporary SERNAC evidence also documents regular weekday and
card-based promotions at both chains: Servicio Nacional del Consumidor,
``SERNAC denuncia publicidad enga\~nosa de farmacias,'' 6~June~$2006$,
\url{https://www.sernac.cl/portal/619/w3-article-864.html}.}

\subsection{The quantity response to an undercut}\label{ssec:desc_undercut_quantity}

I define an undercut event as a transaction-price cut of at
least 15\% from a stable seven-day plateau that makes the initiator newly
cheapest. I compare the initiator's $\log(1+q)$ with the mean $\log(1+q)$ of its two rivals around the event. Here, $q$ is daily quantity sold. I subtract the mean of this difference over the seven days before the cut. Figure~\ref{fig:undercut_quantity}
plots this relative log quantity from seven days before through seven
days after the cut. Before the advertisement ban, the response on the cut date is large but reverses on days one and two. After the ban, the
cut-date response is small and the increase appears on days one and two.
These event paths motivate the demand specification in
Section~\ref{sec:demand_estimation}, which allows the gain from belonging to the
low-price group to differ across the two regimes for the same medicine.

\begin{figure}[htbp]
\centering
\caption{Relative quantity around an undercut}
\label{fig:undercut_quantity}
\includegraphics[width=0.88\linewidth]{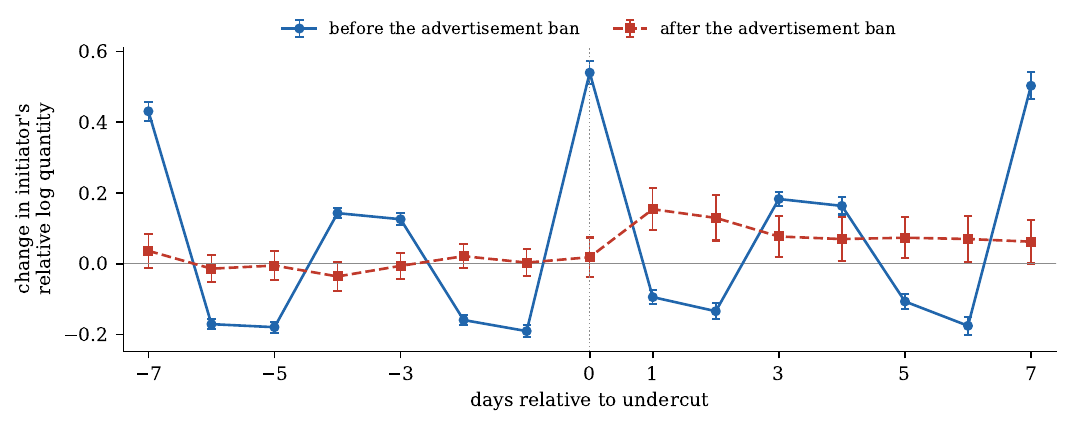}
\par\smallskip\flushleft
{\footnotesize \textbf{Note}: Pre-ban: 4,947 events through 31~August~2007; post-ban: 373 events from
6~November~2007. Relative log quantity is normalized to its days $-7$ to $-1$
mean. Bars show 95\% confidence intervals with standard errors clustered by
medicine. The demand-estimation post-ban sample starts on 12~November.}
\end{figure}
\FloatBarrier

\subsection{Organizing coordinated price increases}\label{ssec:desc_organization}

The three chains' daily transaction-price paths show that observed increase
attempts became more concentrated by laboratory and date before the first
successful three-chain implementation.
I group attempts out of Tier~0 when they concern medicines supplied by the same laboratory
on the same date. From \eventnum{$2$~September} through \eventnum{$5$~November~$2007$},
the \eventnum{$25$} unsuccessful attempts form \eventnum{$24$} such groups
across \eventnum{$22$} dates. From \eventnum{$6$~November} through
\eventnum{$8$~December}, \eventnum{$42$} unsuccessful attempts form
\eventnum{$28$} laboratory--date groups.\footnote{Online Appendix Section~\ref{app:lab_batches}, ``Laboratory batches,'' reports both windows.}
Table~\ref{tab:lab_batches} reports the composition and outcomes from
6~November onward. This
increase in grouping by laboratory and date disciplines the laboratory review
process in Section~\ref{sec:mechanisms}.
Figure~\ref{fig:fail_rate_time} places these attempts in the full
episode. An unsuccessful attempt is an increase not matched by both rivals. The initiator may revert or remain at the higher price.

\begin{figure}[!htbp]
\centering
\caption{Coordination events over the episode}
\label{fig:fail_rate_time}
\begin{subfigure}[t]{0.84\linewidth}
  \centering
  \caption{Events by tier transition}
  \label{fig:fail_rate_time_a}
  \includegraphics[width=\linewidth]{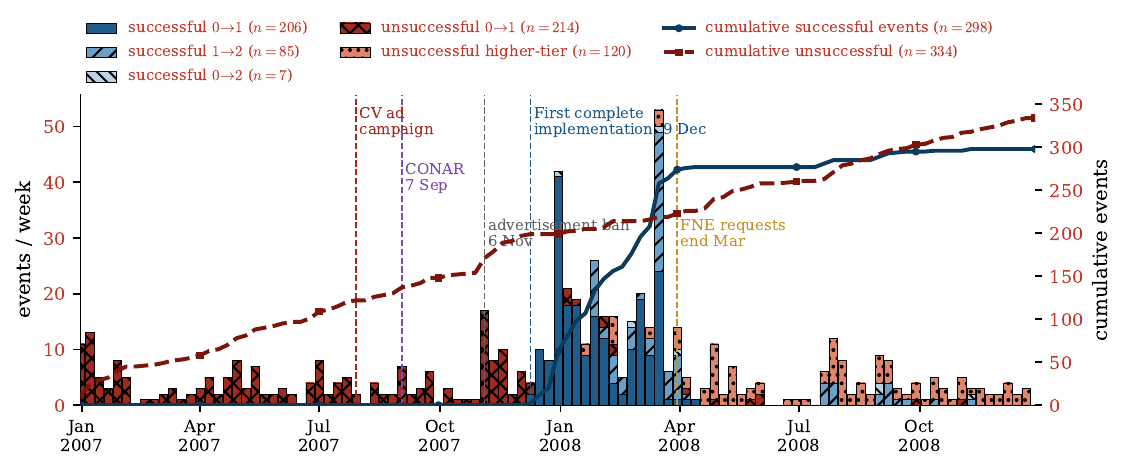}
\end{subfigure}

\medskip
\begin{subfigure}[t]{0.84\linewidth}
  \centering
  \caption{Initiations by chain and period}
  \label{fig:fail_rate_time_b}
  \includegraphics[width=\linewidth]{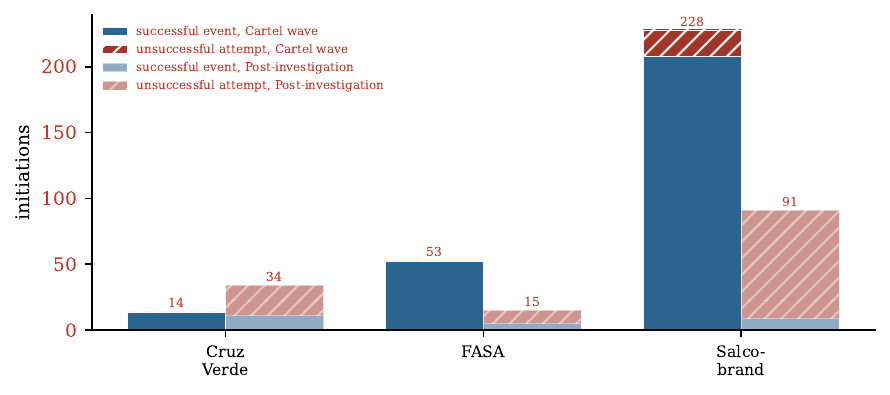}
\end{subfigure}
\par\smallskip\flushleft
{\footnotesize \textbf{Note}: Panel~(a): daily-price events, 31~December~2006--3~January~2009.
Panel~(b): cartel wave (9~December~2007--31~March~2008) versus
1~April--31~December~2008. Each event is one medicine-level
initiation attempt; Panel~(b) assigns it to the initiating chain. Bar labels
count successful and unsuccessful initiations combined. ``Unsuccessful
higher-tier'' combines Tier~1 and Tier~2 attempts. CV denotes Cruz Verde.}
\end{figure}
\FloatBarrier

Panel~(\subref{fig:fail_rate_time_a}) of
Figure~\ref{fig:fail_rate_time} shows repeated unsuccessful attempts before the
first complete implementation on 9~December~$2007$. Successful Tier~$0\to1$ events
then spread through March~$2008$, while Tier~$1\to2$ events began on 27~January~$2008$. A few further increases succeeded in April after the FNE's
requests reached the chains. After the last of these April increases, successful
increases stopped until late July, although unsuccessful attempts continued.\footnote{Online Appendix Section~\ref{app:post_investigation_attempts}, ``After the investigation,'' separates these attempts from cuts away from an established coordinated price.} These later events describe the continuation of coordination beyond the dynamic estimation period.

Panel~(\subref{fig:fail_rate_time_b}) shows that Salcobrand
initiated 208 of the 273 successful events during the cartel wave. Initiation
was less concentrated from 1~April through 31~December~$2008$. Salcobrand initiated 9
successful events, Cruz~Verde 11, and FASA 5.\footnote{Online Appendix Section~\ref{app:additional_descriptive}, ``Leadership by chain and period,'' reports the full decomposition.}

After the FNE investigation began, Cruz Verde and FASA shifted from emails and telephone calls to face-to-face contact. Salcobrand retained its ordinary communication channels.\footnote{Online
Appendix~\ref{app:direct_evidence}, ``(p)~Communication after the investigation began,'' reports that Cruz~Verde and FASA moved from emails
and telephone calls to face-to-face contact, while Salcobrand continued to use
its ordinary channels \citep[consid.~88]{tdlc2012sentencia}.}
The later price pattern retained some laboratory grouping. Three of the 19 Tier~$1\to2$ events, excluding direct Tier~$0\to2$ events, after 31~March~$2008$ occurred on
25~July~$2008$ for Pfizer. This formed one same-day, same-laboratory cluster.\footnote{Online Appendix Section~\ref{app:post_investigation_attempts}, ``After the investigation,'' reports the construction.}

To characterize which medicines entered coordination earlier, I regress the
calendar date of each drug's first successful event out of Tier~0 on its
pre-ban price coefficient, a drug-specific descriptive
demand-slope measure constructed from pre-ban price and quantity variation,
mean pre-ban revenue, mean positive gap between wholesale cost
and price, and drug type (chronic or acute crossed with prescription or
over-the-counter status). These characteristics
explain \eventnum{$10.6\%$} of the variation in these event dates. Adding
laboratory fixed effects raises the $R^2$ to \eventnum{$63.6\%$}
(Table~\ref{tab:sequence_reg}), so
the observed ordering is strongly associated with laboratory
membership. Predetermined
product attributes explain little of the remaining order within laboratories.\footnote{See Online Appendix Section~\ref{app:within_laboratory_order}, ``Within-laboratory coordination order,'' Table~\ref{tab:sequence_within}.}

Across the 36 laboratories, the correlation between mean log
pre-ban product revenue and mean date of the first successful event out of
Tier~0 is $-0.22$. Laboratories with higher-revenue medicines entered earlier
on average.\footnote{The correlation
aggregates 210 coordinated medicines to the laboratory level. Drug-level
pre-ban revenue is mean daily chain--drug revenue before the advertisement ban;
timing is the laboratory mean of each medicine's first successful event date
out of Tier~0.}

\begin{table}[ht]
\centering
\caption{Laboratory membership and timing of the first successful event out of Tier~0}
\label{tab:sequence_reg}
\begin{minipage}{0.95\linewidth}
\centering
\small
\begin{tabularx}{\linewidth}{@{}>{\raggedright\arraybackslash}Xc@{}}
\toprule
Specification (\eventnum{$N=210$}) & $R^{2}$ \\
\midrule
Price coefficient, pre-ban revenue, below-cost gap, and drug-type fixed effects
    & \eventnum{$0.106$} \\
\quad $+$ laboratory fixed effects
    & \eventnum{$\mathbf{0.636}$} \\
\bottomrule
\end{tabularx}
\par\smallskip\flushleft
{\footnotesize \textbf{Note}: The outcome is the calendar day of
each medicine's first successful event out of Tier~0. The sample contains the 210 medicines whose first successful event out of Tier~0
occurred by 30~June~$2008$. The pre-ban price coefficient is a drug-specific descriptive
demand-slope measure; Section~\ref{sec:demand_estimation} estimates the demand
system jointly across medicines. Pre-ban revenue and the below-cost gap are
measured in thousands of CLP.}
\end{minipage}
\end{table}

\FloatBarrier

\subsection{The movement beyond restored margins}\label{ssec:desc_second_round}

Table~\ref{tab:desc_tier_markups} reports quantity-and-cost-weighted markups of $9$--$12\%$ below cost in Tier~0, $22$--$25\%$ above cost in Tier~1, and $43$--$48\%$ above cost in Tier~2.
Figure~\ref{fig:cost_price} compares two series. The first is the cross-medicine mean of
Salcobrand's wholesale costs for the 221 medicines with observed costs. The second is
the quantity-weighted transaction price across three chains and 222 medicines.
From the common 1~November~$2007$ base
to 1~May~$2008$, the retail-price index rose $43\%$ while the wholesale-cost
index rose $2.2\%$.\footnote{Online Appendix Section~\ref{app:below_cost_losses}, ``Monthly losses on below-cost medicine sales,'' reports the associated descriptive peso shortfall and its cost-data limitation.}

\begin{figure}[ht]
\centering
\caption{Wholesale cost and retail price}
\label{fig:cost_price}
\includegraphics[width=0.92\linewidth]{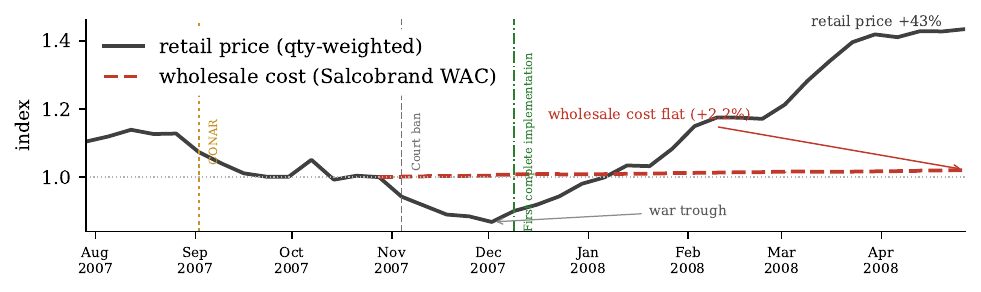}
\par\smallskip\flushleft
{\footnotesize \textbf{Note}: Both indices equal one on 1~November~2007. Dashed: wholesale cost;
solid: retail price. WAC denotes wholesale acquisition cost; ``Court ban''
denotes the advertisement ban. Samples and weighting are described in the text.}
\end{figure}

\begin{table}[ht]
\centering\small
\caption{Cost-relative markups by tier and chain}
\label{tab:desc_tier_markups}
\begin{minipage}{0.95\linewidth}
\centering
\begin{tabularx}{\linewidth}{@{}>{\raggedright\arraybackslash}X c c c@{}}
\toprule
                    & Tier~0 & Tier~1 & Tier~2 \\
                    & Below cost & Margin restoration & Rent extraction \\
\midrule
Cruz Verde          & $-0.101$ & $0.236$  & $0.454$ \\
FASA                & $-0.117$ & $0.220$  & $0.435$ \\
Salcobrand          & $-0.089$ & $0.249$  & $0.476$ \\
\bottomrule
\end{tabularx}
\par\smallskip\flushleft
{\footnotesize \textbf{Note}: Each cell is a
quantity-times-cost-weighted average of the cost-relative markup:
$\sum_j Q_{ij}(p^{z}_j-c_j)/\sum_j Q_{ij}c_j$. Here $p^{z}_j$ is medicine
$j$'s fixed benchmark transaction price in Tier~$z$, for $z\in\{0,1,2\}$, $c_j$ is its wholesale cost,
and $Q_{ij}$ is chain~$i$'s total quantity sold across the full demand panel,
using the same quantity weights for all three tiers.
The fixed benchmark tier prices and costs are common across chains within a medicine, so chain differences in this table come from quantity weights. The sample contains 222 medicines, each contributing
once to each chain--tier cell. Online Appendix Section~\ref{app:event_coding},
\textquotedblleft Construction of price tiers\textquotedblright, gives the tier construction and imputations.}
\end{minipage}
\end{table}

\FloatBarrier

Coordination continued after the Tier~$0\to1$ movement had restored
positive margins. From 31~December~$2006$ through 3~January~$2009$, the
three chains' daily transaction-price paths record 206 successful Tier~$0\to1$ events, 85 successful
Tier~$1\to2$ events, and seven direct Tier~$0\to2$ events. These categories are mutually exclusive. A direct Tier~$0\to2$ event crosses both adjacent tier boundaries. The first
Tier~$1\to2$ crossing occurred on 27~January~$2008$, 49 days after the first
Tier~$0\to1$ crossings on 9~December~$2007$. Over the full observation window, Salcobrand initiated 59 successful
Tier~$1\to2$ events.\footnote{Online Appendix Section~\ref{app:additional_descriptive}, \textquotedblleft Leadership by chain and period\textquotedblright, Table~\ref{tab:desc_chain} reports successful events by initiator and the rates of movement between adjacent tiers.}

Margin restoration proceeded faster even after accounting for how many
medicines were eligible for each movement and how long they remained eligible.
I compare the 65 days beginning with the first success for each movement. The Tier~$0\to1$ window is
9~December~$2007$--11~February~$2008$. The Tier~$1\to2$ window is
27~January--31~March~$2008$.
For Tier~$0\to1$, medicines remain eligible until their first successful increase
out of Tier~0. For Tier~$1\to2$, they enter the eligible group the day after
completing Tier~$0\to1$ and leave after completing Tier~$1\to2$.
I divide the number of first transitions by the total time these medicines
remain eligible. The resulting rates are 9.83 and 5.39 transitions per
100 medicine-weeks, respectively. Margin restoration occurred at 1.83 times
the rate of rent extraction. Direct Tier~$0\to2$ events are counted separately
because they skip the waiting time in Tier~1.\footnote{Online Appendix Section~\ref{app:additional_descriptive}, ``Leadership by chain and period,'' Table~\ref{tab:desc_chain}, Panel~B, reports the calculation.}

Section~\ref{sec:mechanisms} compares the observed diffusion paths with separately
estimated models in which the leader's subjective weight either updates after
rent-extraction attempts or remains fixed at one.

\FloatBarrier
\endgroup

\begingroup
\section{Demand}\label{sec:demand_estimation}

I estimate a demand system to measure how price differences and recent price
changes affect purchases at each chain. These estimates translate the price
changes described in Section~\ref{sec:descriptive} into changes in quantity
and profit. Section~\ref{sec:mechanisms} then uses these profits to estimate
the dynamic pricing game.
\begingroup
\subsection{Dynamic instrumental-variable (IV) demand}\label{ssec:demand_model}

I estimate a simple-logit demand system that allows purchases to respond to
current prices, low-price-group membership, and recent price changes. The
low-price-group term captures the additional demand a chain attracts by pricing
sufficiently below its rivals. The price-history terms allow a recent increase
or cut to affect demand beyond its effect on the current price. They also capture an
additional response when a chain raises its price and remains alone at the high
price. These features connect demand to the price changes documented in
Section~\ref{sec:descriptive}.

I use a daily panel of recorded transaction prices and quantities. Let
$j\in\{1,\ldots,J\}$ index medicines, $i\in\mathcal I$ index the three pharmacy
chains, and $t$ index days. For each medicine, consumers choose among the three
chains and the outside good of not purchasing from any of them. I retain a
medicine--day only when all three chains report positive prices and quantities.

Let $q_{ijt}$ denote quantity and $N_j$ the market size for medicine $j$. I set
$N_j$ equal to the maximum total daily quantity sold by the three chains on a
day with complete observations, divided by $0.92$, plus one unit.\footnote{The
maximum accommodates daily sales fluctuations. The divisor uses the three
chains' combined market share reported in the Introduction, and the additional
unit ensures a positive outside share.}
The inside and outside shares are therefore
\[
s_{ijt}=\frac{q_{ijt}}{N_j},\qquad
s_{0jt}=1-\sum_{i\in\mathcal I}s_{ijt}.
\]

Consumer $n$'s utility from purchasing medicine $j$ at chain $i$ on day $t$ is
$u_{nijt}=\delta_{ijt}+\varepsilon_{nijt}$, where $\delta_{ijt}$ is mean utility
and $\varepsilon_{nijt}$ is an idiosyncratic taste shock. I normalize the outside
good's mean utility to zero, so $u_{n0jt}=\varepsilon_{n0jt}$. With i.i.d.
Type I extreme-value taste shocks, the logit shares imply
$\log(s_{ijt}/s_{0jt})=\delta_{ijt}$.

Mean utility depends on the transaction price $p_{ijt}$, low-price-group
membership $L_{ijt}$, recent increases $A^+_{ijt}$, recent cuts $A^-_{ijt}$, and
the unmatched high-price position $H_{ijt}$. I allow the price coefficient and
the low-price-group coefficient to differ before and after the advertisement
ban, while keeping the three price-history coefficients common across periods.
Let $\mathrm{Post}_t$ equal one in the post-ban estimation window and zero in
the pre-ban window. The estimating equation is
\begingroup\small
\setlength{\jot}{2pt}
\begin{equation}
\begin{split}
\log(s_{ijt}/s_{0jt})={}&\phi_{ij}+\zeta_t
-\alpha_{\mathrm{pre}}p_{ijt}(1-\mathrm{Post}_t)
-\alpha_{\mathrm{post}}p_{ijt}\mathrm{Post}_t\\
&+\lambda_{\mathrm{pre}}L_{ijt}(1-\mathrm{Post}_t)
+\lambda_{\mathrm{post}}L_{ijt}\mathrm{Post}_t\\
&+\gamma_{\mathrm{inc}}A^+_{ijt}+\gamma_{\mathrm{cut}}A^-_{ijt}
+\gamma_H H_{ijt}+\xi_{ijt}.
\end{split}
\label{eq:demand_estimating}
\end{equation}
\endgroup
Here $\phi_{ij}$ and $\zeta_t$ are chain--medicine and common-day fixed effects,
and $\xi_{ijt}$ is the remaining demand shock. Price is measured in thousands
of Chilean pesos. The positive price coefficients $\alpha_{\mathrm{pre}}$ and
$\alpha_{\mathrm{post}}$ enter with a minus sign. The $\lambda$ coefficients
measure the additional mean utility associated with low-price-group membership.

A chain belongs to the low-price group when its price is at least five percent
below the highest chain price. Writing
$p^{\mathrm{max}}_{jt}=\max_{k\in\mathcal I}p_{kjt}$, I define
\[
L_{ijt}=\mathbf 1\!\left\{p_{ijt}<p^{\mathrm{max}}_{jt}\ \text{and}\
\frac{p^{\mathrm{max}}_{jt}-p_{ijt}}{p^{\mathrm{max}}_{jt}}\geq0.05\right\}.
\]
Up to two chains can therefore belong to the low-price group for the same
medicine on the same day.

The recent-increase and recent-cut variables accumulate price changes over the
current day and the preceding six calendar days. For consecutive daily prices,
let $\Delta p_{ijt}=\log(p_{ijt}/p_{ij,t-1})$, and set it to zero when
consecutive daily prices are unavailable. With $\ell$ denoting the number of
days before $t$,
\[
A^+_{ijt}=\sum_{\ell=0}^{6}\max\{\Delta p_{ij,t-\ell},0\},\qquad
A^-_{ijt}=\sum_{\ell=0}^{6}\max\{-\Delta p_{ij,t-\ell},0\}.
\]

Finally, $H_{ijt}$ captures a recent increase that leaves the chain alone at a
sufficiently high price. Let
$\mathrm{gap}_{ijt}=\max\{0,\log p_{ijt}-\log\max_{k\ne i}p_{kjt}\}$
when $\max_{k\ne i}p_{kjt}\leq0.95p_{ijt}$, and zero otherwise. The threshold
measures the highest rival's discount relative to the chain's own price. I
define $H_{ijt}=\mathrm{gap}_{ijt}\mathbf 1\{A^+_{ijt}>0\}$. This term becomes
zero when a rival matches the price, even if the recent increase still
contributes to $A^+_{ijt}$. Unlike $L_{ijt}$, the three price-history variables
measure magnitudes rather than binary events. I use these same definitions in
both demand regimes and recompute them under counterfactual price paths using
only prices available through day $t$.

\endgroup

\begingroup
\subsection{Identification and estimation}\label{ssec:demand_identification}\label{ssec:demand_est}

I use national listed prices as instruments for transaction prices.
A listed price is the chain's centrally chosen national
pre-discount schedule, whereas transaction prices incorporate the discounts and
promotions applied to customers \citep[consids.~27--31]{tdlc2012sentencia}.
The case record also describes list prices and manufacturers' suggested prices
as strategic benchmarks \citep[para.~96]{fne2008requerimiento}. Transaction
prices can respond to transaction-stage demand shocks through discounts and
promotions, making them endogenous. My maintained identifying assumption is
that changes in the national listed-price schedule, including the schedule
against which weekly discounts are set, are predetermined strategic pricing
choices and do not respond to the residual consumer-demand shock $\xi_{ijt}$.
Conditional on chain--medicine and day fixed effects and the other demand
controls, listed prices affect purchases through transaction prices. The
institutional timing of the national schedule therefore motivates the exclusion
restriction. I interact both prices with the pre- and post-ban indicators to estimate
the two price coefficients.

I assume that the two low-price-group interactions and the recent price-history
variables $A^+_{ijt}$, $A^-_{ijt}$, and $H_{ijt}$ are uncorrelated with the
remaining demand shock after accounting for chain--medicine and day fixed
effects. These fixed effects absorb persistent differences in demand across
chain--medicine pairs and common daily shocks. I estimate all seven demand
coefficients jointly by two-stage least squares and cluster standard errors by
medicine.\footnote{The first-stage partial $R^2$ is $0.953$
before the ban and $0.965$ afterward; the corresponding medicine-clustered
$F$-like statistics for the excluded instruments are $7{,}962$ and $9{,}448$.}

The estimation sample covers 1~January~2006--2~September~2007 before the
advertisement ban and 12~November~2007--31~December~2008 afterward. I exclude
the intervening period, which covers the advertising rulings and extends
through the first calendar week of the binding ban. Listed prices are
available for 589,624 chain--medicine--day observations, or 95.5 percent of the
eligible demand sample.

Table~\ref{tab:demand_hetero} reports the demand estimates. The price
coefficient changes little after the advertisement ban, falling from $0.034$
to $0.028$. The difference is $-0.006$, with a standard error of $0.003$, while
the mean own-price elasticity changes from $-0.353$ to $-0.314$. These estimates
indicate a modest change in consumers' response to transaction prices.

The gain from belonging to the low-price group falls much more sharply. Its
coefficient declines from $0.207$ before the ban to $0.062$ afterward, a
reduction of 70 percent. The difference is $-0.145$, with a standard error of
$0.018$. Holding current prices, recent price histories, and fixed effects
constant, low-price-group membership raises purchase odds relative to the
outside good by 23.0 percent before the ban and 6.4 percent afterward. Thus,
the main change in estimated demand is the smaller additional gain from
offering a sufficiently low price.

Recent price changes also affect demand after controlling for current prices
and low-price-group membership. The recent-increase coefficient is $-0.138$
(standard error $0.057$). The recent-cut coefficient is $-0.040$ (standard
error $0.047$). It provides no clear evidence of an additional response to recent
cuts. A recent increase carries a further demand loss when the chain remains
alone at the high price. The coefficient on $H$ is $-1.013$ (standard error
$0.101$). This additional loss grows with the log price gap between the chain
and its highest-priced rival.

\begin{table}[t]
\centering\small
\caption{IV demand estimates by regime}
\label{tab:demand_hetero}
\begin{minipage}{0.95\linewidth}
\centering
\begin{threeparttable}
\begin{tabular*}{\linewidth}{@{\extracolsep{\fill}}lrr@{}}
\toprule
& Pre-ban & Post-ban \\
\midrule
\multicolumn{3}{l}{\textit{Panel A. Regime-specific demand responses}} \\
Price coefficient $\hat\alpha$
& $0.034^{***}$ & $0.028^{***}$ \\
& $(0.006)$ & $(0.005)$ \\
\addlinespace
Low-price-group gain $\hat\lambda$
& $0.207^{***}$ & $0.062^{***}$ \\
& $(0.016)$ & $(0.014)$ \\
\addlinespace
Mean own-price elasticity
& $-0.353$ & $-0.314$ \\
\midrule
\multicolumn{3}{l}{\textit{Panel B. Common price-path responses}} \\
Seven-day cumulative price-increase magnitude $\hat\gamma_{\mathrm{inc}}$
& \multicolumn{2}{c}{$-0.138^{**}$} \\
& \multicolumn{2}{c}{$(0.057)$} \\
Seven-day cumulative price-decrease magnitude $\hat\gamma_{\mathrm{cut}}$
& \multicolumn{2}{c}{$-0.040$} \\
& \multicolumn{2}{c}{$(0.047)$} \\
Unmatched high-price-gap magnitude $\hat\gamma_H$
& \multicolumn{2}{c}{$-1.013^{***}$} \\
& \multicolumn{2}{c}{$(0.101)$} \\
\bottomrule
\end{tabular*}
\begin{tablenotes}[flushleft]\footnotesize
\item \textit{Notes.} Standard errors clustered by medicine are
in parentheses. The Panel~B coefficients multiply unscaled log price-change
and log price-gap magnitudes. $^{*}$, $^{**}$, and $^{***}$ denote significance at the 10,
5, and 1 percent levels. Mean own-price elasticity is the unweighted mean of
$-\hat\alpha_d p_{ijt}(1-s_{ijt})$ evaluated at observed prices and shares in
regime $d$.
\end{tablenotes}
\end{threeparttable}
\end{minipage}
\end{table}

\endgroup

\begingroup
\subsection{Demand-implied quantity and profit effects}\label{ssec:dynamic_demand}

A price change affects profit through both sales of the medicine and complementary basket profit associated with the resulting change in medicine demand. I use the demand
estimates to calculate these two effects for an unmatched price increase and
a unilateral five-percent cut. I treat each unit of medicine demand as one
customer-equivalent unit and let $\mu$ denote complementary basket profit per customer-equivalent unit. Figure~\ref{fig:solo_raise_profit}
shows how the profitability of each action varies with $\mu$ under pre- and
post-ban demand.

I use price changes available in Section~\ref{sec:mechanisms}.
The first increase starts from the observed initial price vector and raises
one chain to Tier~1; I include only chains whose initial price is below Tier~1.
The second starts at Tier~1 and raises one chain to Tier~2. Rivals hold their
prices for seven days, and the unmatched raiser returns to its starting price
next week. The cut is the first ordinary cut: one chain lowers its own initial
price by five percent for seven days. Next week, all three chains charge
95 percent of their respective initial prices, the model's initial price-war
state. Initial prices need not be equal across chains.\footnote{Online
Appendix Section~\ref{app:event_coding}, ``Construction of price tiers,''
reports the price construction.}

The figure evaluates each increase or cut along the fixed two-week price path described above. This fixed path isolates how the estimated demand system changes the payoff from each action.\par
The calculation uses Section~\ref{ssec:mech_payoffs}'s
one-off payoff. It includes seven days at the action prices plus the discounted
seven-day price-history correction caused by the next week's price change.
It excludes continuation values. First-week changes are measured against
holding the starting prices. The next-week correction compares demand at
the next price vector with and without price history; it does not include
the full lasting effect of lower prices. Quantity percentages divide the
first-week change plus this correction by baseline sales over the two weeks,
evaluated at each week's price vector without price-history effects.

Let $a$ denote the increase or cut. Write $\Delta q^0_{ij}(a)$ and
$\Delta\pi^{0,\mathrm{medicine}}_{ij}(a)$ for the first-week changes in daily
quantity and medicine profit, and $\Delta q^1_{ij}(a)$ and
$\Delta\pi^{1,\mathrm{medicine}}_{ij}(a)$ for the second-week effects just
described. Each medicine-profit term is the change in $q(p-c_j)$, where $c_j$
is medicine~$j$'s wholesale cost. Adding profit from other purchases gives
\begin{equation}
\Delta\Pi^{\mathrm{event}}_{ij}(a;\mu)
=7\left[\Delta\pi^{0,\mathrm{medicine}}_{ij}(a)
+\mu\Delta q^0_{ij}(a)\right]
+\beta 7\left[\Delta\pi^{1,\mathrm{medicine}}_{ij}(a)
+\mu\Delta q^1_{ij}(a)\right],
\label{eq:demand_solo_raise_profit}
\end{equation}
where $\beta$ is the weekly discount factor used in
Section~\ref{sec:mechanisms}. Event profit is measured in thousand CLP,
and $\mu$ in thousand CLP per customer-equivalent unit.
The figure varies $\mu$. Its value is estimated in the dynamic game.

\begin{figure}[!htbp]
\centering
\caption{Demand-implied quantity and profit effects of unilateral price changes}
\label{fig:solo_raise_profit}
\begin{subfigure}[t]{0.49\textwidth}
\centering
\caption{Unmatched-increase quantity loss}
\includegraphics[width=\linewidth]{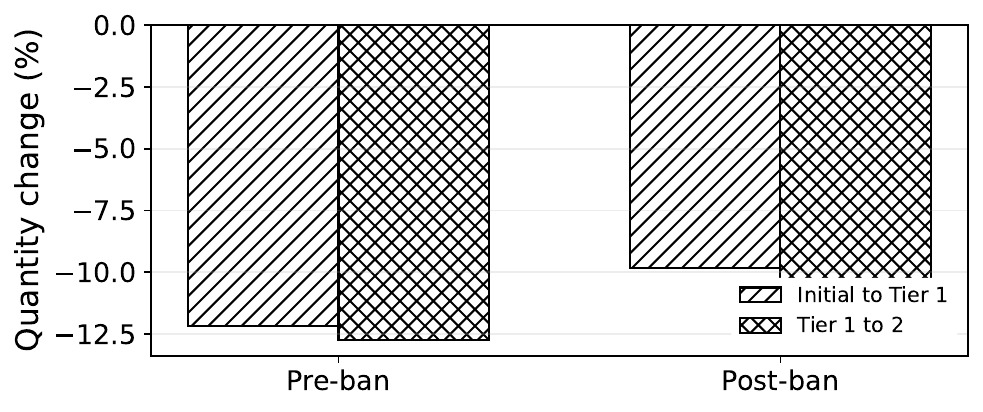}
\end{subfigure}
\hfill
\begin{subfigure}[t]{0.49\textwidth}
\centering
\caption{Initial-cut quantity gain}
\includegraphics[width=\linewidth]{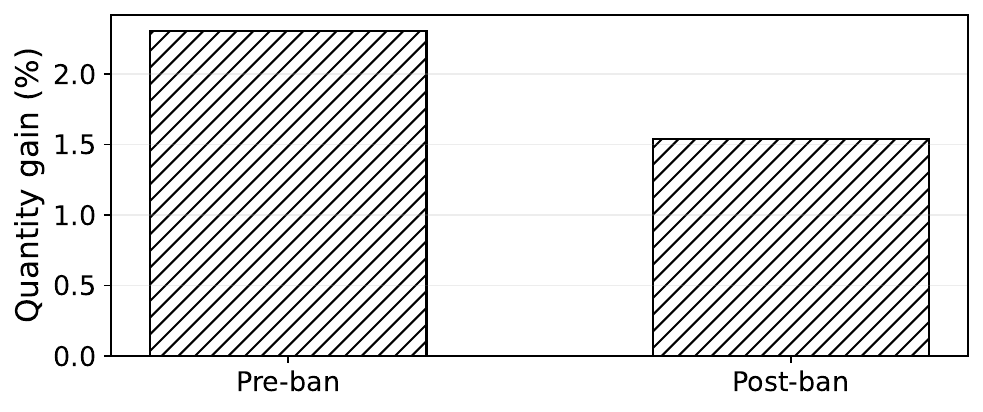}
\end{subfigure}

\begin{subfigure}[t]{0.49\textwidth}
\centering
\caption{Unmatched-increase profit gain/loss}
\includegraphics[width=\linewidth]{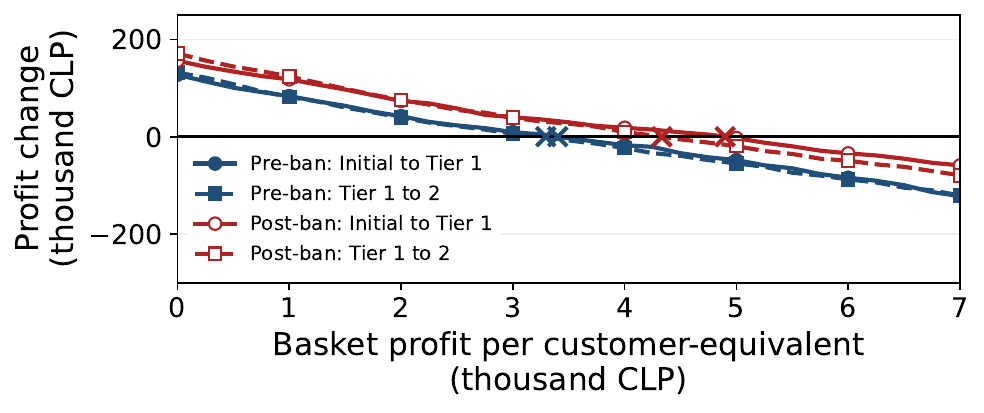}
\end{subfigure}
\hfill
\begin{subfigure}[t]{0.49\textwidth}
\centering
\caption{Initial-cut profit gain/loss}
\includegraphics[width=\linewidth]{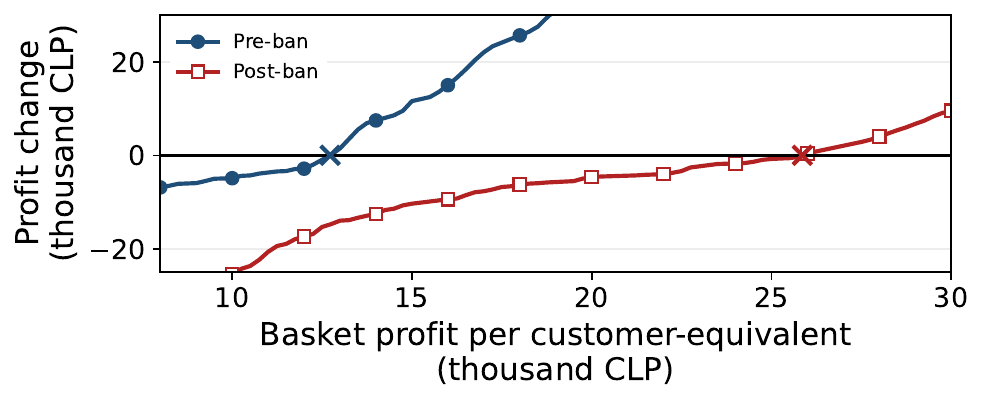}
\end{subfigure}
\begin{minipage}{0.96\linewidth}
\footnotesize\vspace{0.4em}
\textit{Notes.} Medians across medicine--chain pairs: 467 for initial-to-Tier~1 increases;
666 for Tier~$1\to2$ increases and cuts. Quantity changes are percentages of
the two-week baseline; event profits are in thousand CLP. Complementary basket profit
$\mu$ is in thousand CLP per customer-equivalent unit. Calculations follow
Equation~\eqref{eq:demand_solo_raise_profit}.
\end{minipage}
\end{figure}

\FloatBarrier

Panel~A reports a median quantity loss of
$12.2$ percent before the ban and $9.8$ percent afterward for an
eligible initial-to-Tier~1 increase. The Tier~$1\to2$ losses are
$12.8$ and $11.2$ percent. Panel~B reports a median gain of
$2.3$ percent before the ban and $1.5$ percent afterward for the same
initial five-percent cut. All these percentages use the two-week event denominator.\footnote{For a separate seven-day comparison of predicted and observed losses following unmatched price increases, see Online Appendix Section~\ref{app:observed_quantity_responses}, ``Observed outcomes around price changes.''}

Panel~C shows how lost basket profit changes the return to an
unmatched increase. The initial-to-Tier~1 profit curve crosses zero at
$\mu=3.40$ thousand CLP before the ban and $4.90$ afterward;
the Tier~$1\to2$ crossings are $3.29$ and $4.33$.
Panel~D shows the opposite tradeoff for a cut. Additional basket profit can
cover the lost medicine margin. The initial-cut curve crosses zero at
$\mu=12.73$ thousand CLP before the ban and $25.86$ afterward.
These are one-off payoff comparisons, not predictions of the optimal action.
In Section~\ref{sec:mechanisms}, cutting after verification also starts
the price-war continuation. That future loss can deter a cut even when its
one-off payoff is positive.
\footnote{Online Appendix Section~\ref{app:flow_payoff_thresholds},
``Daily quantity and profit,'' reports profit thresholds for the price-change
week, excluding next-week price-history effects and later continuation values.
Section~\ref{app:seven_day_quantity_losses}, ``Demand specifications and quantity effects of model actions,''
compares event quantity losses from increases and gains from cuts across
demand specifications. Online Appendix
Table~\ref{tab:demand_cutoff_robustness} shows that the post-ban cutoff is
higher across alternative price-history definitions.}

\endgroup
\FloatBarrier

\FloatBarrier
\endgroup

\begingroup
\section{Dynamic Pricing Game}\label{sec:mechanisms}

\begingroup
This section develops a dynamic pricing game to model the
transition from price war to coordinated price increase. It organizes the
transition around three mechanisms. First, the advertisement-ban mechanism operates through the
estimated demand system: the post-ban regime has a smaller fitted reward from being in the low-price group and thereby changes the payoff from undercutting. Second, the
procedure-verification mechanism allows initially uninformed followers
 to learn a repeatable one-two-three price-increase procedure, which supports coordinated margin restoration across
medicines (Tier~$0\to1$). Third, after verification, the model
allows a candidate leader's subjective weight on sequential
follower choices, relative to raising alone, to respond to earlier
rent-extraction outcomes (Tier~$1\to2$).
The \textbf{Adaptive Confidence model} updates this weight after each complete or
incomplete selected rent-extraction proposal using a Beta-shaped outcome-updating rule. In this model, actual followers make their own sequential choices. The
separately estimated \textbf{Unit Confidence model} instead fixes the leader's
subjective weight at one after verification, retaining the same sequential follower choices, and provides the benchmark.

The demand estimates determine quantities, which enter flow payoffs through medicine margins and complementary basket profit. The same $\mu$ measures complementary basket profit in every block. What changes after the advertisement ban is the demand generated by a low price. Observed laboratory dates and counts determine scheduled reviews in Game~2. Verification changes what firms know about the price-increase procedure, while followers continue to make their own participation decisions. Both specifications use the same demand estimates and review rules but estimate their economic parameters separately. Margin restoration can follow once the procedure is verified. Rent extraction goes further and requires the leader to assess whether rivals will participate.

\endgroup

\subsection{Environment, states, timing, and flow payoffs}
\label{ssec:mech_setup}\label{ssec:mech_payoffs}\label{ssec:mech_complete_setup}

\begingroup
\paragraph{Model setup.}
Let $i\in\mathcal I=\{\mathrm{CV},\mathrm{FASA},\mathrm{SB}\}$ index pharmacy
chains, $j\in\mathcal J=\{1,\ldots,J\}$ index medicines, and
$t\in\mathcal T^{\mathrm{obs}}=\{52,\ldots,117\}$ index Sunday--Saturday model
weeks. Here $t$ indexes weeks. In the demand section it indexes days. Thus, $\mathcal I$ contains Cruz~Verde, FASA, and Salcobrand,
$J=222$, and $\mathcal T^{\mathrm{obs}}$ runs from
31~December~2006--6~January~2007 ($t=52$) through
30~March--5~April~2008 ($t=117$).

I divide this period into two games because the binding advertisement ban marks
an exogenous regime change in the model. It changes both payoffs and how firms
receive opportunities to raise prices. The
decentralized price-adjustment game (Game~1) covers
31~December~2006--3~November~2007 ($t=52,\ldots,95$). During this game, each
eligible medicine may receive a weekly price review.
Within Game~1, the model switches to calibrated campaign cut depths and a separately estimated review frequency in
5--11~August~2007 ($t=83$). The earlier block ends with
29~July--4~August~2007 ($t=82$), and the campaign block ends with
28~October--3~November~2007 ($t=95$). I use 5~August as the model cutoff.%
\footnote{The record dates the campaign only to August~2007. The model uses
5~August~2007 because it is the first complete model week beginning in August;
it is not an observed campaign-start date.}
The laboratory-mediated coordination game (Game~2) covers
4~November~2007--5~April~2008 ($t=96,\ldots,117$). The binding advertisement ban
took effect on 6~November~2007, during the first week of Game~2. From Game~2
onward, the model uses post-ban payoffs and closes the independent
Game~1 review process; laboratory dates and product counts
determine scheduled reviews in Game~2. The model assigns both changes to the
4--10~November~2007 week
($t=96$).

In the forward simulation, the observed laboratory dates and
medicine counts determine the scheduled reviews. At each date, the model randomly
selects up to the observed count from the eligible medicines without replacement.
The stationary value calculation uses the corresponding average review opportunity. Random assignment and cut-blocked reviews can leave some medicines without an unblocked opportunity to restore margins within the simulation period. I therefore add supplementary Tier~$0\to1$ reviews so that the allocation of review slots does not by itself exclude these medicines from a verified opportunity. These reviews provide an opportunity, not a guaranteed increase. There are no supplementary Tier~$1\to2$ reviews.\footnote{Online Appendix Section~\ref{app:forward_simulation}, ``Forward simulation,'' specifies the timing and eligibility of supplementary reviews.} These reviews determine when a medicine can be considered for an increase. Verification determines whether the followers know the procedure for coordinating that increase.
A candidate leader is a chain facing the choice to raise first. The first chain to accept becomes the leader. Before verification, only Salcobrand faces this choice. After verification, any of the three chains can do so.\footnote{See Online Appendix
Section~\ref{app:lab_batches}, ``Laboratory batches.''}

Before verification, the leader and followers solve different
subjective games under the same post-ban demand system. Salcobrand knows the
laboratory procedure and solves Game~2. Each uninformed follower instead solves
a decentralized game with the structure of Game~1's campaign block (Block~2),
recomputed using post-ban demand. It expects the other chains, including
Salcobrand, to follow that decentralized game and does not anticipate the
laboratory procedure. In each unverified week, the followers become informed
with probability $\eta_0$. Once verified, they switch permanently to Game~2
and, in both specifications, choose whether to follow according to their own incentives.
In Adaptive Confidence, a candidate leader evaluates a
rent-extraction proposal by weighting the value of sequential follower choices
against the value of raising alone. Complete and incomplete attempts update
this subjective weight. Under the Unit Confidence model, the candidate leader is
assigned a subjective weight of one on the sequential follower outcome law after verification; each follower still chooses whether to join.

For drug $j$, let
$z_{jt}\in\{I,0^{(0)},\ldots,0^{(K)},1,2\}$ denote its price state and
$\mathbf p_j^{z_{jt}}=(p_{ij}^{z_{jt}})_{i\in\mathcal I}$ its persistent prices
at the three chains. State $I$ is the initial-price state, states
$0^{(0)},\ldots,0^{(K)}$ are the price-war levels, and states 1 and 2 are the two
coordinated-price tiers. $K$ is the last index in the price-war grid, which starts at zero. At these prices, consumers choose among the three
chains and the outside good. I define the transition rules after the within-week
timing.

At the start of week $t$, let $G_t=(R_t,S_t^r,F_t^r)$ denote the
public information common across drugs. The superscript $r$ denotes rent
extraction. The indicator $R_t$ records whether the chains have verified the
price-increase procedure. The counts $S_t^r$ and $F_t^r$ record completed and
incomplete rent-extraction attempts made before week $t$.
Equation~\eqref{eq:mech_willingness_belief} maps these counts, the initial subjective weight
$m_0^r$, and prior strength $\nu^r$ (the effective number of outcomes
represented by the prior) into the candidate leader's
subjective weight $m_t^r$ on the value of sequential follower choices relative
to raising alone. This weight is not the probability that both rivals
ultimately follow, since sequential choices can also produce incomplete
participation. Thus,
$m_t^r$ is derived from
$G_t$ rather than included as a separate state variable. Because $G_t$ is common
across drugs, verification or the outcome of an attempt for one drug can affect
later choices for another. For a drug outside punishment, I write the economic state as
\begin{equation}
x_{jt}=(z_{jt},G_t).
\label{eq:mech_state}
\end{equation}
Punishment is a separate continuation, with value
$V^{\mathrm{pun}}_{ij,t}(z_{jt})$, in which the drug has no future increase
opportunities. Whenever a transition enters punishment, this value replaces
the ordinary continuation value $V_{ij,t}(x_{jt})$; the same price state outside punishment retains its increase
opportunities. Section~\ref{ssec:mech_stage} defines the punishment transitions.
Calendar time $t$ determines which game and policy apply, but it
is not part of the economic state.\footnote{I solve each block as a stationary infinite-horizon game. The value
calculation approximates learning on a grid; the simulation keeps the learning
counts and chain-specific prices. Online Appendix Sections~\ref{app:backward_solution}
and~\ref{app:forward_simulation} explain the solution and simulation, respectively.}

\paragraph{Weekly timing.}
For medicines outside punishment, price decisions occur at reviews,
except that in Game~2 before verification, followers can also cut outside
reviews. The simulation conditions these additional cuts on their observed
weekly totals.\footnote{Online Appendix Section~\ref{app:forward_simulation}, ``Forward simulation,'' describes how these totals enter the simulation.} Punished medicines follow their cut-only continuation
each week, without waiting for a review.
Game~1 draws each eligible product with a common weekly probability. Game~2
uses the laboratory dates and their product counts. At a review, the three
chains first choose cut or hold simultaneously. Any cut lowers the drug's price
state in the following week and ends its decisions for the current week. If all
three chains hold, the increase stage follows. Game~1 uses simultaneous
increase-or-hold choices. Game~2 uses the candidate-leader and, when applicable,
follower choices. Incomplete participation changes current prices only.
complete participation advances the drug's persistent price tier. A cut after
verification instead starts the punishment continuation and removes the drug
from future increase opportunities. I place every price action on the first day
of its model week. The model does not use the action's calendar day within the
week.

\begingroup
Figure~\ref{fig:weekly_timing} summarizes the within-week sequence.

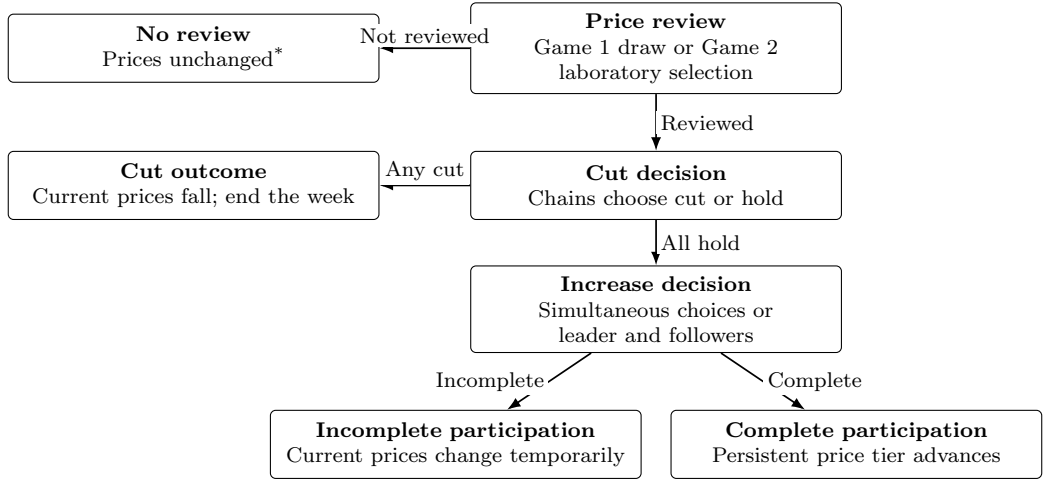
\begin{figure}[htbp]
\centering
\caption{Within-week timing of the dynamic game}
\label{fig:weekly_timing}
\begin{tikzpicture}[
  flowbox/.style={draw=black, text=black, rounded corners=2pt, align=center,
                 minimum height=0.9cm, text width=0.28\linewidth,
                 inner sep=4pt, font=\scriptsize},
  arrow/.style={-{Latex[length=1.8mm]}, line width=0.65pt,
                draw=black},
  branchlabel/.style={font=\scriptsize, fill=white, inner sep=1.5pt}
]
\node[flowbox] (review) {\textbf{Price review}\\Game~1 draw or Game~2 laboratory selection};
\node[flowbox, below=0.75cm of review] (cutstage) {\textbf{Cut decision}\\Chains choose cut or hold};
\node[flowbox, below=0.60cm of cutstage] (choices) {\textbf{Increase decision}\\Simultaneous choices or\\leader and followers};

\node[flowbox, left=1.2cm of review] (noopp) {\textbf{No review}\\Prices unchanged\textsuperscript{*}};
\node[flowbox, left=1.2cm of cutstage] (cutout) {\textbf{Cut outcome}\\Current prices fall; end the week};
\node[flowbox, below=0.75cm of choices, xshift=-2.65cm] (incomplete) {\textbf{Incomplete participation}\\Current prices change temporarily};
\node[flowbox, below=0.75cm of choices, xshift=2.65cm] (complete) {\textbf{Complete participation}\\Persistent price tier advances};

\draw[arrow] (review.west) -- node[branchlabel, above] {Not reviewed} (noopp.east);
\draw[arrow] (review) -- node[branchlabel, right] {Reviewed} (cutstage);
\draw[arrow] (cutstage.west) -- node[branchlabel, above] {Any cut} (cutout.east);
\draw[arrow] (cutstage) -- node[branchlabel, right] {All hold} (choices);
\draw[arrow] (choices) -- node[branchlabel, left] {Incomplete} (incomplete);
\draw[arrow] (choices) -- node[branchlabel, right] {Complete} (complete);
\end{tikzpicture}
\par\smallskip\flushleft
{\footnotesize \textbf{Note}: Both games follow this sequence. \textsuperscript{*}Before verification in
Game~2, followers can also cut without a review; these cuts are conditioned
on observed weekly totals. A cut after verification starts punishment and
ends future increase opportunities.}
\end{figure}
\endgroup

\FloatBarrier
\paragraph{Flow payoffs.}
The post-ban demand estimates imply a smaller extra gain from offering a low price, while ordinary price sensitivity changes little. This reduces the customer-attraction benefit of undercutting in the flow payoffs below.

Let $\mathbf p_j^{\mathrm{old}}\equiv\mathbf p_j^{z_{jt}}$ be the opening price
vector. Let $\chi_{jt}$ denote the specified proportional cut
depth for drug $j$ in the current block. A cut action selects the corresponding
lower price, so its depth enters the feasible action prices as well as the
next-week state. Block~1 uses the ordinary five-percent cut. Block~2 allows
drug-specific cut depths calibrated to observed price paths, and the post-ban games use the expanded cut-price
space.  The weekly choice maps it into one of three action-week price scenarios.
If all chains hold at both the cut and increase stages,
$a^c=a^u=\mathbf 0$ and the price vector remains
$\mathbf p_j^{\mathrm{old}}$. If at least one chain cuts, the binary profile
$a^c$ identifies the cutters and the price vector is $\mathbf p_j^{a^c}$. The
week ends at that point, so $a^u=\mathbf 0$. Otherwise, the binary profile $a^u$
identifies the raisers and the price vector is $\mathbf p_j^{a^u}$. Each profile
has one entry per chain, equal to one for a cutter or raiser. I denote the
resulting scenario-specific vector by $\mathbf p_j^{\mathrm{new}}(a^c,a^u)$,
or simply $\mathbf p_j^{\mathrm{new}}$.

Demand estimates from Section~\ref{sec:demand_estimation} enter
the dynamic game as fixed inputs and are not re-estimated. For a transition
from $\mathbf p_j^{\mathrm{old}}$ to $\mathbf p_j^{\mathrm{new}}$, I construct
$A^+$, $A^-$, and $H$ from the opening-day price change and the new cross-chain
price gaps using the seven-day definitions that follow
Equation~\eqref{eq:demand_estimating}. Because no further price change occurs
within the week, the opening-day change remains in the seven-day demand history
on all seven model days.
In the scenario formulas, I suppress the day index on $A^+,A^-$ and $H$;
they are evaluated at the displayed old and new price vectors.
Let $g(t)\in\{\mathrm{ordinary},\mathrm{campaign},
\mathrm{laboratory}\}$ index the three model environments defined above.
The demand index satisfies $d(\mathrm{ordinary})=d(\mathrm{campaign})=\mathrm{pre}$
and $d(\mathrm{laboratory})=\mathrm{post}$. Thus, $g(t)$ identifies the model
environment and $d(g(t))$ selects its demand coefficients.
The fitted mean utility and the daily quantity it implies are
\begingroup\small
\setlength{\jot}{2pt}
\begin{equation}
\label{eq:unilateral_diversion}
\begin{aligned}
\widehat\delta^{g(t)}_{ij}
(\mathbf p_j^{\mathrm{new}},\mathbf p_j^{\mathrm{old}})
={}&\widehat\phi_{ij}
-\widehat\alpha_{d(g(t))}p_{ij}^{\mathrm{new}}
+\widehat\lambda_{d(g(t))}
L_{ij}(\mathbf p_j^{\mathrm{new}})
+\widehat\gamma_{\mathrm{inc}}A^+_{ij}
+\widehat\gamma_{\mathrm{cut}}A^-_{ij}
+\widehat\gamma_HH_{ij},\\[2pt]
\widetilde q^{g(t)}_{ij}
(\mathbf p_j^{\mathrm{new}},\mathbf p_j^{\mathrm{old}})
={}&N_j\frac{\exp\{\widehat\delta^{g(t)}_{ij}
(\mathbf p_j^{\mathrm{new}},\mathbf p_j^{\mathrm{old}})\}}
{1+\sum_{k\in\mathcal I}\exp\{\widehat\delta^{g(t)}_{kj}
(\mathbf p_j^{\mathrm{new}},\mathbf p_j^{\mathrm{old}})\}}.
\end{aligned}
\end{equation}
\endgroup
All demand objects follow
the definitions in Section~\ref{sec:demand_estimation}. Coefficients are
reported in Table~\ref{tab:demand_hetero}. When evaluating an action, the game
sets the common day effect $\zeta_t$ and residual demand shock $\xi_{ijt}$ to zero
because they capture observed daily variation rather than systematic demand at
the scenario prices. If prices do not change, $\mathbf p_j^{\mathrm{new}}=
\mathbf p_j^{\mathrm{old}}$ and $A^+=A^-=H=0$.

\begingroup

\begin{samepage}
To compare quantities across drugs, I normalize daily quantity by
$\widetilde N\equiv\operatorname{median}_j N_j$. The drug-specific cost $c_j$
is the median wholesale-cost measure. It is time invariant and common across
chains. The $\mu$ is complementary basket profit per customer-equivalent unit, measured in thousand CLP.\footnote{Online Appendix Section~\ref{app:below_cost_losses}, ``Monthly
losses on below-cost medicine sales,'' gives the cost construction.}
For a given $(a^c,a^u,x_{jt};\mu)$, normalized daily profit
under its action-week price profile is
$\overline\pi_{ij,t}\equiv
[\widetilde q^{g(t)}_{ij}(\mathbf p_j^{\mathrm{new}},
\mathbf p_j^{\mathrm{old}})/\widetilde N]
(p_{ij}^{\mathrm{new}}-c_j+\mu)$.
\end{samepage}

A complete profile becomes the next price tier.\footnote{Online
Appendix Section~\ref{app:event_coding}, ``Construction of price tiers,'' describes the tier-price construction.} An incomplete profile
applies only for the current week; at the opening of the next week, all chains
return to the old price tier.

Let $\Delta\overline\pi_{ij,t+1}$ denote the change in normalized daily profit
next week caused by a price change at the opening of that week. It compares
profit at the same next-week prices with and without that price change in the
seven-day demand history. The difference can be positive or negative. Within each
stationary Bellman block, both terms use that block's fixed demand regime.
Both $\overline\pi_{ij,t}$ and
$\Delta\overline\pi_{ij,t+1}$ are evaluated for the
$(a^c,a^u,x_{jt};\mu)$ on the left-hand side below. Chain $i$'s weekly payoff
is

\begin{samepage}
\begin{equation}
\pi^{g(t)}_{ij}(a^c,a^u,x_{jt};\mu)
=7\overline\pi_{ij,t}+7\beta\Delta\overline\pi_{ij,t+1}.
\label{eq:mech_payoff}
\end{equation}
\end{samepage}

The current-week term applies for seven days whether a raise is complete or
incomplete. A leader's profit nevertheless differs across the two outcomes:
only an incomplete profile can leave the leader uniquely high priced, which is
captured by $H$ in Equation~\eqref{eq:unilateral_diversion}. The observed
two-day median response time is a descriptive timing moment and does not
shorten the model week.

If any chain cuts, the public price state moves to the destination
specified by the applicable cut depth, and all chains charge the resulting
lower price profile at the opening of the next week.
These next-week price changes determine
$\Delta\overline\pi_{ij,t+1}$. The term is zero when the price profile does not
change at the opening of next week. I include this difference for one week and
assume that any price gap is then closed. The event-profit
change $\Delta\Pi^{\mathrm{event}}_{ij}(a;\mu)$ in
Equation~\eqref{eq:demand_solo_raise_profit}, divided by $\widetilde N$,
is the action-minus-hold difference between the corresponding normalized
weekly payoffs in Equation~\eqref{eq:mech_payoff}, when evaluated at the
same price profiles and demand inputs.
\endgroup

\endgroup

\subsection{State transitions and public information}
\label{ssec:mech_stage}\label{ssec:mech_learning}

\paragraph{States and updating.}
The price state $z_{jt}$ indexes the persistent three-chain price
vector $\mathbf p_j^{z_{jt}}$. State $I$ indexes the observed vector in the first model week. A first cut
moves the drug to the initial price-war state $0^{(0)}$. The
ordinary cut moves one five-percent grid step, $0^{(w)}\to0^{(w+1)}$;
a deeper cut can cross several grid steps. The price grid must contain the
destinations implied by the specified cut depths.
States~1 and~2 are the constructed Tier~1 and Tier~2
price vectors described in Section~\ref{sec:background_data}.\footnote{Online
Appendix Section~\ref{app:event_coding}, ``Construction of price tiers,''
distinguishes observed benchmark prices from imputed prices and describes
the Tier~2 prediction method.} No further raise is
available from Tier~2.%
\footnote{Four drugs attain sustained post-Oficio prices above their Tier~2
benchmarks. Online Appendix Section~\ref{app:higher_price_levels}, ``Observed price
levels above Tier~2,'' documents these observations.}

\begingroup
Here $\mathbf p_j^{0^{(w)}}$ is the persistent price vector
$\mathbf p_j^{z_{jt}}$ when $z_{jt}=0^{(w)}$. For the initial price-war state,
each component of $\mathbf p_j^{0^{(0)}}$ is 95 percent of the chain's price in state $I$. Deeper price-war states
satisfy $\mathbf p_j^{0^{(w)}}=0.95^{w}\mathbf p_j^{0^{(0)}}$ for
$w=0,\ldots,K$.

For the ordinary cut depth, the downward-state mapping is $\mathcal D(I)=0^{(0)}$,
$\mathcal D(0^{(w)})=0^{(w+1)}$ for $w=0,\ldots,K-1$, and
$\mathcal D(0^{(K)})=0^{(K)}$. Thus $\mathcal D$ is the ordinary-depth special case of the
drug-week mapping $\mathcal D_{jt}$ defined below. The upward-state mapping is
$\mathcal U(z)=1$ for $z\in\{I,0^{(0)},\ldots,0^{(K)}\}$ and
$\mathcal U(1)=2$. Thus $\mathcal U$ maps a margin-restoration increase
(Tier~$0\to1$) to state~1 and a rent-extraction increase (Tier~$1\to2$) to
state~2. The post-verification cut transition additionally starts the
punishment continuation and is specified separately below.
\endgroup

\begingroup

\paragraph{Price-state transitions.}
In the Bellman calculation, the next-week grid state follows the within-week
sequence in Figure~\ref{fig:weekly_timing}. When a drug is reviewed, the chains
first choose whether to cut. If at least one chain cuts, the drug moves down
the price-war ladder. Write $\mathcal D_{jt}$ for the downward
mapping at cut depth $\chi_{jt}$. It is the ordinary one-step mapping in
Block~1 and incorporates deeper cut destinations in Block~2 and the post-ban
games. The same mapping determines action payoffs, Bellman continuation values,
and forward price transitions. If all three chains hold, they proceed to the
increase stage, where only a complete three-chain increase moves the state up:
\begingroup\small
\setlength{\jot}{2pt}
\begin{equation}
z_{j,t+1}=\begin{cases}
\mathcal D_{jt}(z_{jt}),&\text{the drug is reviewed and at least one chain cuts},\\
\mathcal U(z_{jt}),&\text{the drug is reviewed, no chain cuts, and the increase is complete},\\
z_{jt},&\text{otherwise}.
\end{cases}
\label{eq:mech_war_transition}
\end{equation}
\endgroup
The cut branch applies except when a drug not already in
punishment cuts with $R_t=1$ at the opening of the week: that deviation starts
punishment at $0^{(0)}$ next week; subsequent punishment cuts follow
$\mathcal D_{jt}$, while verification at the close of a week does not punish
that week's cut.
The final case includes no review, an all-hold profile, and an incomplete
increase. The identity and number of cutters affect current flow payoffs through the
cut-price vector $\mathbf p_j^{a^c}$, but they do not otherwise affect the
transition.

\begingroup
The week-opening indicator $R_t$ records whether the followers
know the common procedure. In Game~2, each week with $R_t=0$ has verification
probability $\eta_0$: $\Pr(R_{t+1}=1\mid R_t=0)=\eta_0$. Otherwise the followers
remain uninformed and continue to play their post-ban-demand version of
Game~1 Block~2. Verification is common across drugs, is permanent, and does
not itself change any drug's price state. Choices during week $t$ use $R_t$;
a verification at its close changes play from week $t+1$. The leader
anticipates this information transition, while uninformed followers do not
include the laboratory procedure in their subjective continuation values.
Game~1 price changes never change $R_t$.
\endgroup

\paragraph{Rent extraction participation and belief updating.}
No outcomes update beliefs in Game~1 or before verification
in Game~2. After verification in Game~2, only the outcomes
of Tier~$1\to2$ attempts update beliefs. Each initiated Tier~$1\to2$ proposal produces one
public outcome. Complete participation advances the drug's persistent state
from 1 to 2 and counts as a success. An incomplete attempt leaves the state at
1, counts as a failure, and leaves the product eligible for later laboratory
reviews.
All products selected in week $t$ use the same opening counts, and their outcomes
are aggregated when the week closes. Let $n_{S,t}^r$ and $n_{F,t}^r$ denote the
numbers of complete and incomplete selected attempts in that week. The count
state evolves as
\begin{equation}
 (S_{t+1}^r,F_{t+1}^r)
 =(S_t^r+n_{S,t}^r,F_t^r+n_{F,t}^r).
\label{eq:mech_belief_count_transition}
\end{equation}
If no Tier~$1\to2$ product is selected, both weekly counts are zero. Tier~$0\to1$
outcomes do not enter either count.

At a rent-extraction proposal, the candidate leader's subjective
weight on the sequential-following value is
\begin{equation}
m_t^r=m^r(G_t;m_0^r,\nu^r)
=\frac{\nu^r m_0^r+S_t^r}{\nu^r+S_t^r+F_t^r},
\label{eq:mech_willingness_belief}
\end{equation}
where $m_0^r$ is the initial subjective weight and $\nu^r$ is the prior strength.
The subjective weight $m_t^r$ is derived from the common count state $G_t$ rather than
included as a separate state variable.

\paragraph{Punishment continuation.}
If any chain cuts after verification, the drug enters the punishment
continuation in the following week at the initial price-war state $0^{(0)}$.\footnote{Online Appendix~\ref{app:direct_evidence},
``(m)~Punishment: the threat was a return to the price war,'' reports the
FNE's account of this threat \citep[para.~96, pp.~35--36]{fne2008requerimiento}.
This account motivates the return to price competition; the timing and
transition rules here are model assumptions.}
Raise opportunities are then shut down for that drug. The common cut stage
continues to operate. All three chains holding leaves the price state unchanged. A cut applies $\mathcal D_{jt}$ and $0^{(K)}$ remains absorbing. The
verification state remains one. This drug contributes no further
successes or failures to the public counts, but eligible Tier~$1\to2$ attempts
for other drugs continue to update them. Its value is the separate
$V^{\mathrm{pun}}_{ij,t}(z_{jt})$. Public learning does not reopen its increase
opportunities.

\subsection{Dynamic choices and policy functions}\label{ssec:mech_choices}

\begingroup
\begingroup
\paragraph{When prices are reviewed.}
In Game~1, each eligible drug is reviewed with weekly probability
$\omega_1$ before the campaign and $\omega_2$ during it. Game~2 uses the review
process described in Section~\ref{ssec:mech_setup}. Before verification, only
Tier~$0\to1$ increases are available; afterward, both Tier~$0\to1$ and
Tier~$1\to2$ increases are available.\footnote{Online Appendix Section~\ref{app:forward_simulation}, ``Forward simulation,'' describes the review process.}

\paragraph{Choices at reached nodes.}
At a review, the three chains simultaneously choose whether to cut or hold.
Before verification, each follower evaluates cutting and holding using
Block~2 of Game~1, solved under post-ban demand. It expects the other chains,
including Salcobrand, to play that game and does not interpret a price increase
as a coordinated proposal. The leader anticipates these follower policies but evaluates cutting and
holding using its Game~2 continuation values. After a cut, it can propose a
new Tier~$0\to1$ increase at a later review. After verification, a cut
leads to the punishment continuation in Section~\ref{ssec:mech_stage} and ends
future increase opportunities for that drug. If all three chains hold in
Game~1, they then choose increase or hold simultaneously.

At the Game~2 increase stage, leadership is determined as follows.
Before verification, only Salcobrand chooses between leading and waiting on a
Tier~$0\to1$ increase. It evaluates leading using the followers' Game~1 policies
described above and accounts for the probability of verification in its
next-week continuation value. The followers continue to use those policies
and do not interpret the increase as a coordinated proposal. After verification,
the model assigns equal probability to each of the six possible orders of the
three chains. In that order, each chain chooses between leading and waiting.
The first chain that chooses to lead is the unique leader. If all three wait,
no increase occurs.

\begingroup
A node history $h$ records the current decision stage and acting
chain, the realized chain order when applicable, and all actions already
observed in that week's review. At a candidate-leader node it therefore
identifies the candidate's position and which earlier candidates have waited;
at a follower node it also identifies the leader and any earlier follower
action. Simultaneous choices at the current stage are not yet observed.
The next equation separates three cases. The cases are an increase before verification, a margin-restoration increase after verification, and a rent-extraction increase after verification. Only in the last case does the candidate leader weight the value of sequential follower choices against raising alone.
For candidate leader $i$, the value of leading depends on verification
and the proposed tier transition,
\begingroup\small
\setlength{\jot}{2pt}
\begin{equation}
\bar v^{\mathrm{lead}}_{ij,t}
=\begin{cases}
v^{\mathrm{unverified}}_{ij,t},
& R_t=0,\quad 0\to1,\\[1.5pt]
v^{\mathrm{seq}}_{ij,t},
& R_t=1,\quad 0\to1,\\[1.5pt]
v^{\mathrm{solo}}_{ij,t}
+m_t^r\!\left(v^{\mathrm{seq}}_{ij,t}-v^{\mathrm{solo}}_{ij,t}\right),
& R_t=1,\quad 1\to2.
\end{cases}
\label{eq:mech_proposal_realization}
\end{equation}
\endgroup
All three values are conditional on reaching the increase stage, after all chains have chosen to hold.
Conditional on chain $i$ leading at state $x_{jt}$, these values
are independent of which earlier candidates waited. The sequential branch
averages over the two follower orders, so no node-history index is needed.
Let $a^u$ denote the three chains' increase profile, with $i$ denoting the
candidate leader and $k,k'$ the other two chains. I suppress the known cut
profile in the flow payoff. The three values combine weekly profits with
discounted continuation values,
\begingroup\small
\begin{equation}
\begin{aligned}
v^{\mathrm{unverified}}_{ij,t}
&=\mathbb E\!\Bigl[
\pi^{g(t)}_{ij}(a^u,x_{jt};\mu) +\beta\bigl\{(1-\eta_0)V_{ij,t+1}(x_{j,t+1}^{(0)})
+\eta_0 V_{ij,t+1}(x_{j,t+1}^{(1)})\bigr\} \mid a_i^u=1,x_{jt}\Bigr],\\[2pt]
v^{\mathrm{solo}}_{ij,t}
&=\mathbb E\!\bigl[
\pi^{g(t)}_{ij}(a^u,x_{jt};\mu)
+\beta V_{ij,t+1}(x_{j,t+1}) \mid a_i^u=1,\ a_k^u=0,\ a_{k'}^u=0,\ x_{jt}\bigr],\\[2pt]
v^{\mathrm{seq}}_{ij,t}
&=\mathbb E_{\mathrm{seq}}\!\left[
\pi^{g(t)}_{ij}(a^u,x_{jt};\mu)
+\beta V_{ij,t+1}(x_{j,t+1})
\mid a_i^u=1,x_{jt}\right].
\end{aligned}
\label{eq:mech_lead_value_components}
\end{equation}
\endgroup
The flow payoff $\pi^{g(t)}_{ij}$ is defined in
Equation~\eqref{eq:mech_payoff}, with $g(t)=\mathrm{laboratory}$
and $d(g(t))=\mathrm{post}$ here and the known
cut profile omitted. In the unverified value, the expectation is over the
other two chains' actions under their post-ban Game~1 Block~2 policies.
This value applies only to Salcobrand before verification. The resulting
increase profile determines next week's price state as described in
Section~\ref{ssec:mech_stage}. Conditional on that profile,
$x_{j,t+1}^{(0)}$ is the next-week state if verification does not occur,
with $R_{t+1}=0$, and $x_{j,t+1}^{(1)}$ is the next-week state if it does,
with $R_{t+1}=1$. These outcomes have probabilities $1-\eta_0$ and $\eta_0$,
respectively. Verification changes next week's information and continuation
play, not the current increase profile.
The solo value fixes both rivals' actions at hold. The expectation
$\mathbb E_{\mathrm{seq}}$ uses the followers' sequential Game~2
choices in both specifications, with the first anticipating the second's
decision. Each follower uses Equation~\eqref{eq:mech_ccp}, so participation can
be incomplete at either tier. Unit Confidence fixes $m_t^r=1$ without removing
these follower decision nodes. Before verification, the unverified branch
uses the uninformed followers' policies in both specifications. Continuation values account for subsequent public-state updates,
including learning.

The weight $m_t^r$ affects only the candidate's Tier~$1\to2$ lead value.
it does not change followers' conditional choices. Complete and incomplete
Tier~$1\to2$ attempts update the public counts through
Equation~\eqref{eq:mech_belief_count_transition}.
\endgroup

At node $h$, let $a_i^h\in\{0,1\}$ denote chain $i$'s choice
between taking the relevant action and holding or waiting. The profiles
$a^c(h)$ and $a^u(h)$ collect the cut and increase actions resulting from this
choice and all remaining within-week choices, denoted
$a_{\mathrm{rem}}^h$. These include the other chains' simultaneous choices
at the current stage and every later action, including chain $i$'s own later
actions. Using the information and beliefs described above, the
chain evaluates each action as
\begingroup\small
\setlength{\jot}{2pt}
\begin{equation}
v^h_{ij,t}(a_i^h;x_{jt})=
\mathbb E_{a_{\mathrm{rem}}^h}
\left[
\pi^{g(t)}_{ij}\bigl(a^c(h),a^u(h),x_{jt};\mu\bigr)
+\beta V_{ij,t+1}(x_{j,t+1})
\mathrel{\big|}a_i^h,x_{jt},h
\right],
\label{eq:mech_choice_value}
\end{equation}
\endgroup
At a candidate-leader node, the action value in
Equation~\eqref{eq:mech_choice_value} uses the subjective evaluation in
Equation~\eqref{eq:mech_proposal_realization}:
$v^h_{ij,t}(1;x_{jt})=\bar v^{\mathrm{lead}}_{ij,t}$.
The wait value $v^h_{ij,t}(0;x_{jt})$ integrates over the remaining candidates'
lead-or-wait decisions and any ensuing follower decisions, including those
of chain $i$ if it is subsequently a follower. After verification, if no candidate remains, waiting
leaves all prices unchanged for the week. Before verification, Salcobrand is
the only candidate, so its waiting branch contains only the uninformed
followers' decentralized choices. Both branches include discounted next-week
continuation values.
Conditional on reaching a stochastic decision node $h$, the
binary choice follows the rule below. Verified followers use this same
sequential logit rule in both specifications.
\begin{equation}
\Pr(a_i^h=1\mid x_{jt},h)=
\Lambda\!\left(
\frac{v^h_{ij,t}(1;x_{jt})-v^h_{ij,t}(0;x_{jt})}{\tau}
\right),
\label{eq:mech_ccp}
\end{equation}
$\Lambda$ is the logistic cdf and $\tau$ is a fixed action-scale normalization
described in Online Appendix Section~\ref{app:backward_solution}.

\paragraph{Dynamic value.}
Let $V_{ij,t}(x_{jt})$ denote chain $i$'s continuation value for
drug $j$ outside punishment at the state defined in Equation~\eqref{eq:mech_state}.
This value combines current profits with discounted future payoffs.
Before verification, the leader evaluates future payoffs using Game~2,
whereas the followers expect all three chains to compete as in
Block~2 of Game~1, solved under post-ban demand.
Both face the same demand, but only the leader understands the procedure.
After verification, all three chains evaluate future payoffs using Game~2.
The continuation values satisfy
\begingroup\small
\setlength{\jot}{2pt}
\begin{equation}
V_{ij,t}(x_{jt})=
\mathbb E\left[
\pi^{g(t)}_{ij}(a^c,a^u,x_{jt};\mu)
+\beta V_{ij,t+1}(x_{j,t+1})
\mid x_{jt}
\right]
+\mathcal E_{ij,t,\tau}(x_{jt}),
\label{eq:mech_bellman}
\end{equation}
\endgroup
where the expectation is taken under chain $i$'s information and
beliefs about the other chains' behavior. It covers review opportunities,
action profiles $(a^c,a^u)$, and the resulting state transitions
described in Section~\ref{ssec:mech_stage}.
The model-environment index $g(t)$ and demand index $d(g(t))$
are defined in Section~\ref{ssec:mech_payoffs}. Transitions into punishment
use $V^{\mathrm{pun}}_{ij,t+1}(z_{j,t+1})$ in place of the ordinary continuation.
The term $\mathcal E_{ij,t,\tau}$ denotes
expected logit surplus at the stochastic decision nodes with
action scale $\tau$.\footnote{This is the expected taste shock to the chosen action under mean-zero Type~I extreme-value shocks, weighted by the probability of reaching each stochastic node.}

\endgroup
\endgroup

\subsection{Estimation, identification, and inference}\label{ssec:mech_estimation}

\begingroup
Both specifications condition on observed laboratory dates and
batch capacities.\footnote{Online Appendix Section~\ref{app:forward_simulation}, ``Forward simulation,'' describes the construction of review opportunities.}
\endgroup

\begingroup
I calibrate drug-specific campaign cut depths to observed weekly prices and end-of-campaign price levels, while cut choices remain endogenous. Conditional on this calibration, I estimate each specification separately. Unit Confidence estimates four economic parameters, $\boldsymbol\theta_U=(\mu,\eta_0,\omega_1,\omega_2)$. These are complementary basket profit $\mu$ per customer-equivalent unit, the weekly verification probability while followers remain uninformed $\eta_0$, and weekly review probabilities before and during the Game~1 campaign, $\omega_1$ and $\omega_2$. The same $\mu$ enters both games within each specification. Unit Confidence fixes the subjective weight at one after verification. Adaptive Confidence jointly estimates $\boldsymbol\theta_A=(\mu,\eta_0,\omega_1,\omega_2,m_0^r,\nu^r)$. It adds the initial subjective weight and prior strength for six parameters in total. In the criterion below, $\boldsymbol\theta$ denotes the full parameter vector $\boldsymbol\theta_A$ or $\boldsymbol\theta_U$ for the specification being estimated, with its corresponding search region $\Theta_F$. Follower choices remain endogenous in both specifications.

I set
the annual discount factor to $0.80$, so
$\beta=0.80^{1/52}\approx0.9957$ per week.%
\endgroup

For each week, I construct two cumulative counts. The first counts medicines that have moved beyond Tier~0. The second counts medicines that have reached Tier~2.
In simulation, the Tier~$0\to1$ count records a medicine only when a
laboratory-mediated increase in Game~2 succeeds. A complete independent
increase in Game~1 does not enter this count, even if it advances the
medicine's model price state.
Each medicine enters each count at most once. In the data, a direct
Tier~$0\to2$ event enters both observed boundary counts but remains one observed
price event.

\begingroup
\begingroup
I compare data moments with their simulation means in 9 loss blocks.
Within each block, I average the squared standardized discrepancies. The
criterion then averages the 9 block losses with equal weight.
Let $b=1,\ldots,9$ index blocks and $\upsilon=1,\ldots,n_b$ index the
statistics within block $b$, such as weekly prices or price-state shares.
For a block containing a single count, $n_b=1$.
Write $M^{\mathrm{obs}}_{b\upsilon}$ for the observed statistic,
$M^{\mathrm{sim}}_{b\upsilon,B}(\boldsymbol\theta;\tau)$ for its mean across $B$ simulated paths,
and $\sigma_{b\upsilon}$ for its fixed scale. Define
\begingroup\small
\setlength{\jot}{2pt}
\[
\begin{aligned}
\mathcal Q_{b,B}(\boldsymbol\theta;\tau)
&=\frac{1}{n_b}\sum_{\upsilon=1}^{n_b}
\left[
\frac{M^{\mathrm{sim}}_{b\upsilon,B}(\boldsymbol\theta;\tau)-M^{\mathrm{obs}}_{b\upsilon}}{\sigma_{b\upsilon}}
\right]^2,\\
Q_B(\boldsymbol\theta;\tau)
&=\frac{1}{9}\sum_{b=1}^{9}\mathcal Q_{b,B}(\boldsymbol\theta;\tau).
\end{aligned}
\]
\endgroup
I minimize this criterion over the bounded parameter search region
$\Theta_F$:\footnote{Online Appendix Section~\ref{app:estimation_interface}, ``Estimation and inference,'' describes the parameter search region.}
\[
\widehat{\boldsymbol\theta}_B
=\arg\min_{\boldsymbol\theta\in\Theta_F}
Q_B(\boldsymbol\theta;\tau).
\]
\endgroup

I exclude Game~1 cut-count losses because the observed series contains screened events whereas the simulation records all cut actions. Two cumulative-count blocks compare Tier~$0\to1$ and Tier~$1\to2$
medicine counts at five selected weeks. Three price-path blocks compare weekly
average prices before the campaign, during the campaign, and in Game~2.
The remaining four blocks compare terminal mean price-war depth, the Game~2
boundary price, the Game~2 boundary state distribution, and the
verification-time cumulative distribution function (CDF).
\endgroup

\begingroup
For both observed and simulated price-path moments, I first
average daily prices within each drug--chain pair and week, so temporary price
changes contribute according to their duration. I then take the equally
weighted average across drug--chain pairs, in thousand CLP.
\endgroup

\begingroup
Game~2 begins at the start of week~96, carrying forward the price states
generated in Game~1. For week~96, I compare observed and simulated weekly
average prices. The boundary-state moment compares the simulated end-of-week~96
price-state distribution with the distribution mapped from observed week~96
prices, after the first Game~2 update.
\endgroup

\begingroup
Under the weekly information transition, verification need not
coincide with the first complete laboratory-mediated increase. That observed
increase supplies a timing benchmark rather than a direct observation of the
followers' information. The verification-time moment compares
the share of simulated paths verified by the close of each week
with this benchmark indicator, which is zero before week~101 and one thereafter.
A transition from $R_t=0$ to $R_{t+1}=1$ counts as verification
in week~$t$. Informed play starts in week~$t+1$. Thus the week~101 benchmark
refers to verification by that week's close, not to informed play at its opening.
\endgroup

\begingroup
I choose the parameters to match the simulated price paths and
cumulative numbers of medicines reaching each price tier to their observed
counterparts.
\endgroup

The price path and price-state moments discipline $\mu$, $\omega_1$,
and $\omega_2$. The parameter $\mu$ changes the profit consequence of a price difference. The two review probabilities govern how often a drug can change price
in each Game~1 block. Verification timing and early Tier~$0\to1$ diffusion
discipline $\eta_0$.
The level and evolution of the Tier~$1\to2$ cumulative-count
path also inform the two learning parameters. The initial weight $m_0^r$
affects initiation before rent-extraction outcomes accumulate, while prior
strength $\nu^r$ governs how strongly subsequent outcomes change that weight
in Equation~\eqref{eq:mech_willingness_belief}. Early participation and its
subsequent evolution therefore provide information about the two parameters.
I estimate them jointly with the economic parameters using the full criterion.

\begingroup
For Adaptive Confidence, I estimate the parameters conditional
on a learning transition prepared before the parameter search. For each candidate parameter vector, I solve the continuation problems and
simulate price paths to construct the moments. I use the same
moments for both specifications and hold random numbers fixed across parameter
candidates.\footnote{Online Appendix Section~\ref{app:estimation_interface}, ``Estimation and inference,'' describes the numerical search.}

For inference, I use a laboratory-block bootstrap. Each
replication samples laboratories with replacement, retains medicines and their
price histories within each sampled laboratory, and re-estimates demand and
both dynamic models.
\endgroup
\endgroup

\subsection{Results and mechanism assessment}\label{ssec:mech_robust}

\begingroup
At their separately estimated parameters, the two specifications generate similar
Tier~$0\to1$ paths, while Adaptive Confidence more closely matches the observed
cumulative Tier~$1\to2$ path.

Table~\ref{tab:dynamic_estimates_fit} reports separately estimated economic
parameters for Adaptive Confidence and Unit Confidence. A common basket-profit parameter enters both games within each specification. Adaptive Confidence jointly estimates the four economic parameters, the initial subjective weight and prior strength. Unit Confidence estimates only the four economic parameters.

\begin{table}[htbp]
\centering
\caption{Conditional dynamic-game parameter estimates}
\label{tab:dynamic_estimates_fit}
\small
\begin{tabularx}{\textwidth}{>{\raggedright\arraybackslash}Xcc}
\toprule
Parameter & Adaptive Confidence & Unit Confidence \\
\midrule
Complementary basket profit $\mu$ (thousand CLP per customer-equivalent unit) & \shortstack{15.820\\(0.539)} & \shortstack{16.328\\(0.590)} \\
Pre-campaign review probability $\omega_1$ (weekly) & \shortstack{0.02344\\(0.00105)} & \shortstack{0.02328\\(0.00113)} \\
Campaign review probability $\omega_2$ (weekly) & \shortstack{0.10375\\(0.00837)} & \shortstack{0.10500\\(0.00797)} \\
Verification probability $\eta_0$ (weekly, given $R_t=0$) & \shortstack{0.19839\\(0.02045)} & \shortstack{0.11776\\(0.02076)} \\
\midrule
Initial subjective weight $m_0^r$ (estimated) & \shortstack{0.000725\\(0.02011)} & -- \\
Prior strength $\nu^r$ (effective prior count; estimated) & \shortstack{375\\(90.560)} & -- \\
\bottomrule
\end{tabularx}
\begin{minipage}{0.96\textwidth}
\footnotesize\textit{Notes:} Estimates use 100 simulation paths, the observed
laboratory schedule, and fixed calibrated campaign cut depths. Dashes mark
learning parameters that do not apply to Unit Confidence. The estimates use approximate game solutions and, for Adaptive
Confidence, a learning transition held fixed during estimation. Online Appendix
Section~\ref{app:estimation_interface}, ``Estimation and inference,'' reports
the numerical procedure. Parentheses report standard errors from the completed
100-replication paired laboratory-block bootstrap.
\end{minipage}
\end{table}

\begingroup\widowpenalty=10000
Figure~\ref{fig:dynamic_transitions} compares the two estimated specifications
with the observed cumulative tier-transition counts. By week~117, Adaptive
Confidence produces 191.58 medicines for Tier~$0\to1$ and 67.67 for
Tier~$1\to2$, against 208 and 70 in the data. Unit Confidence produces 182.61
and 106.22. Both models begin Tier~$0\to1$ increases too early. Adaptive Confidence more
closely matches the second increase path, whereas Unit Confidence generates
more Tier~$1\to2$ transitions early in Game~2. Because each model has its own estimated economic parameters, the difference between their paths cannot be attributed to the subjective-weight rule alone.
\par\endgroup

\begin{figure}[htbp]
\centering
\caption{Cumulative price-state transitions}
\label{fig:dynamic_transitions}
\begin{subfigure}[t]{0.49\textwidth}
\centering\caption{Tier $0\to1$}
\includegraphics[width=\linewidth]{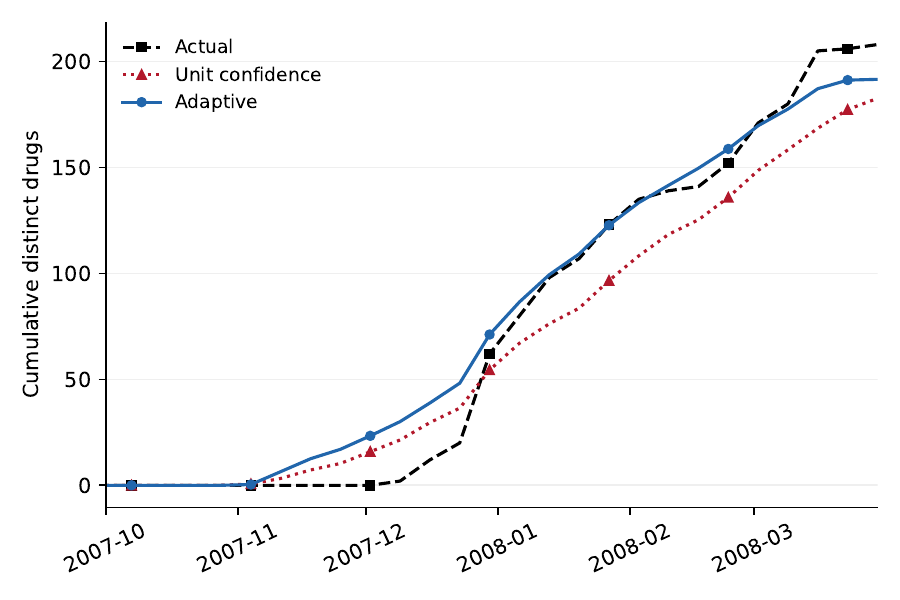}
\end{subfigure}\hfill
\begin{subfigure}[t]{0.49\textwidth}
\centering\caption{Tier $1\to2$}
\includegraphics[width=\linewidth]{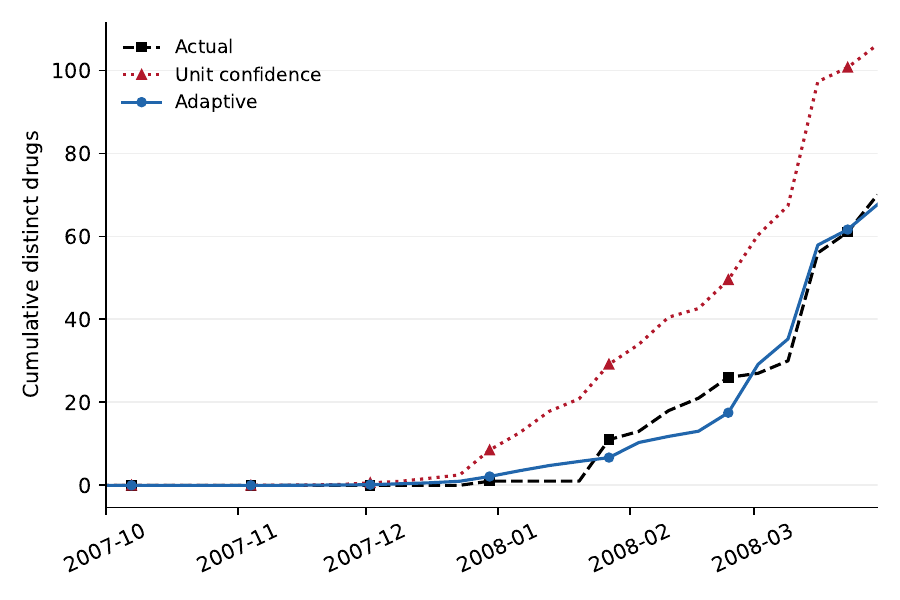}
\end{subfigure}
\begin{minipage}{0.96\textwidth}
\footnotesize\textit{Notes.} Each medicine counts once per tier boundary. Lines average 100 simulated
paths. Black squares: data; red triangles: Unit Confidence; blue circles:
Adaptive Confidence. Simulation starts in January~2007; the displayed window
starts in October~2007.
\end{minipage}
\end{figure}

\FloatBarrier
Figure~\ref{fig:dynamic_cut_price_fit} compares weekly cutting and transaction prices. Both models overpredict the week~117 mean price. The predictions are 14.045 thousand CLP under Adaptive Confidence and 13.978 under Unit Confidence, versus 13.732 in the data. Simulated cuts and screened observed events differ in coverage.

\begin{figure}[!htbp]
\centering
\caption{Cut timing and average transaction prices}
\label{fig:dynamic_cut_price_fit}
\begin{subfigure}[t]{0.49\linewidth}
\centering
\caption{Screened observed events and simulated cut actions}
\label{fig:dynamic_cut_probability}
\includegraphics[width=\linewidth]{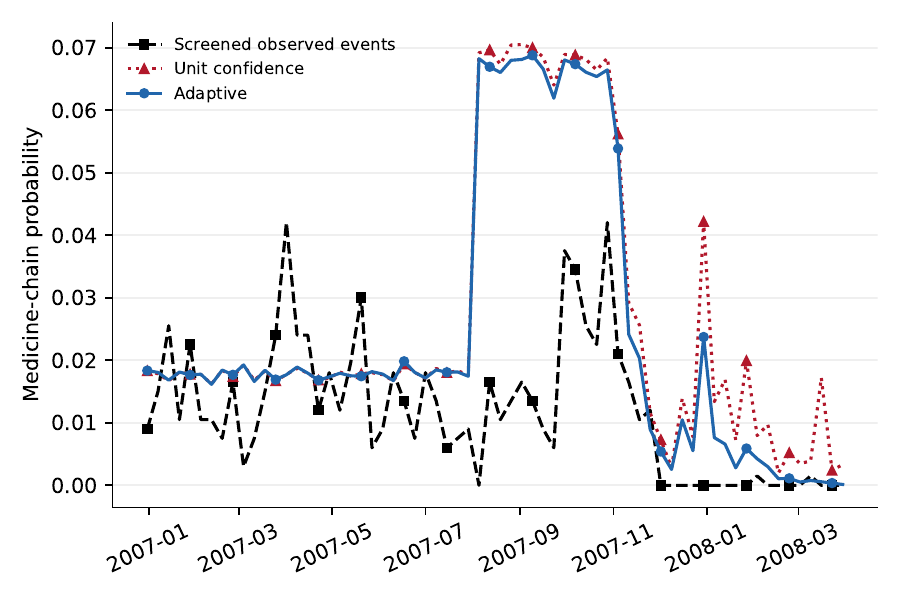}
\end{subfigure}\hfill
\begin{subfigure}[t]{0.49\linewidth}
\centering
\caption{Average transaction price}
\label{fig:dynamic_weighted_price}
\includegraphics[width=\linewidth]{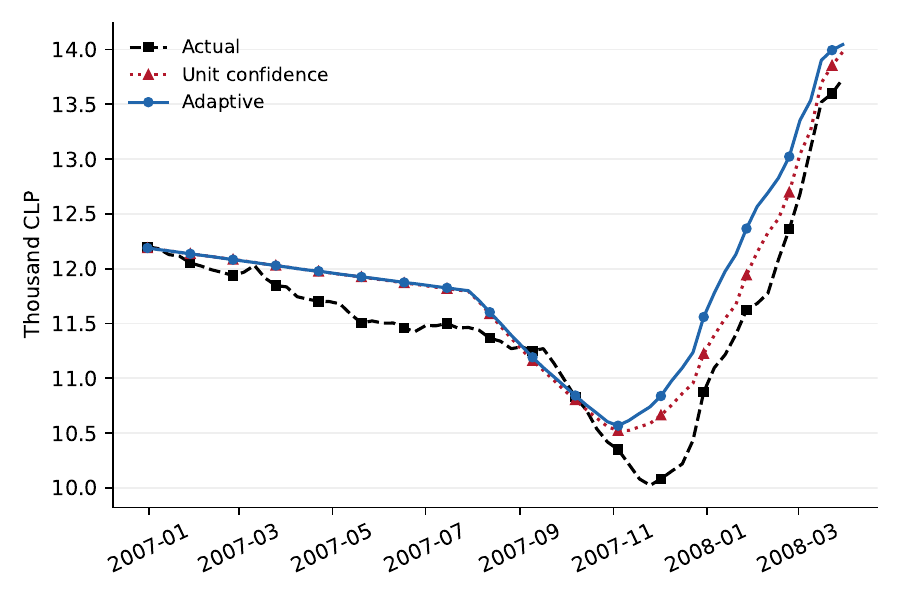}
\end{subfigure}
\medskip
\begin{minipage}{0.96\textwidth}
\footnotesize\textit{Notes.} Panel~(a) compares the timing of all simulated cut actions with screened observed cut events; each series is divided
by 666 medicine--chain cells, but their levels are not directly comparable because their coverage differs. Panel~(b): equally weighted pair-level weekly
mean prices, in thousand CLP. Lines average 100 paths; markers follow
Figure~\ref{fig:dynamic_transitions}.
\end{minipage}
\end{figure}
Figure~\ref{fig:dynamic_initiation_belief} varies the subjective
weight while holding the economic parameters and approximate learning
transition at the Adaptive Confidence estimates in
Table~\ref{tab:dynamic_estimates_fit}. It reports the probability that at least
one candidate initiates a Tier~$1\to2$ increase at a verified review with no cut.

\begin{figure}[htbp]
\centering
\caption{Initiation when the subjective weight is low}
\label{fig:dynamic_initiation_belief}
\includegraphics[width=0.45\textwidth]{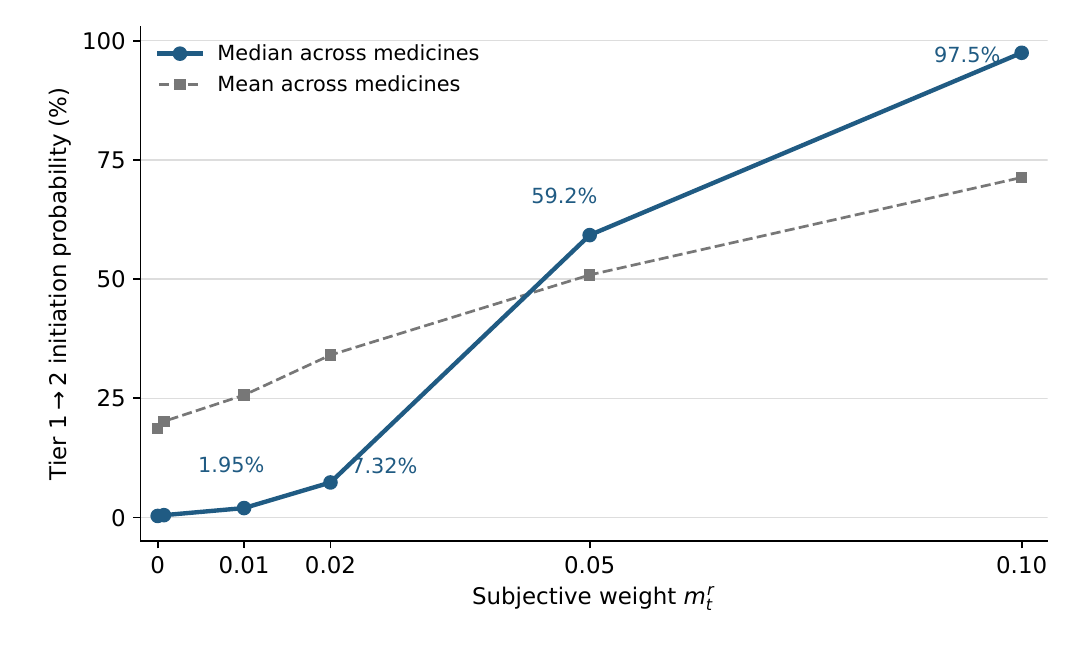}
\begin{minipage}{0.96\textwidth}
\footnotesize\textit{Notes.} Mean and median initiation probabilities across 222 equally weighted
medicines, conditional on a verified review with no cut. Markers are solved
points for $m_t^r\leq0.10$; connecting lines are visual guides.
\end{minipage}
\end{figure}
\FloatBarrier

Figure~\ref{fig:dynamic_initiation_belief} shows a sharp increase in initiation probability as the subjective weight rises from 0.01 to 0.05. The median initiation probability is only
1.95 percent at $m_t^r=0.01$, but rises to 7.32 percent at 0.02 and
59.2 percent at 0.05. In Equation~\eqref{eq:mech_proposal_realization}, the
leader weighs the gain from sequential follower participation relative to
raising alone. When that gain is large, even a small weight can make leading
attractive. The weight changes the leader's valuation. It is not the actual
probability that the followers join.

\endgroup

\FloatBarrier

\FloatBarrier\endgroup

\begingroup
\section{Conclusion}\label{sec:conclusion}

\begingroup

I study how Chile's three largest pharmacy chains moved from a price war to collusion. The demand estimates show that the extra gain from offering a low price fell after the advertisement ban, while the response to ordinary price differences changed little. The ban weakened the incentive to use loss-leader medicines to attract customers, and the sustained price war ended at the same time. The chains then used upstream suppliers to coordinate a sequential price-increase procedure.

I distinguish verification of this procedure from uncertainty about rivals' willingness to participate in further increases. In the model, verification makes the procedure known to the followers. The court record of failed attempts followed by successful repetition motivates this information assumption. Repeated failed attempts preceded the first successful coordinated increase. After that first success, the chains quickly repeated the procedure across medicines to restore positive margins. They subsequently used the same procedure for rent extraction, but these further increases spread more slowly. When considering a rent-extraction increase, the leader still risked losing customers if its rivals did not follow. Knowing how to carry out the procedure did not settle whether the followers would find another increase profitable.

I compare Adaptive Confidence, which updates the weight the leader places on the expected profit from sequential follower responses relative to raising alone, with Unit Confidence, which fixes that weight at one. Complete rent-extraction attempts raise the weight and incomplete attempts lower it. Followers choose whether to join according to their incentives in both models. Adaptive Confidence more closely matches the observed cumulative spread of rent extraction. By the end of the estimation period, it predicts that about 68 medicines reach the rent-extraction level, compared with 70 in the data. Knowing how to coordinate a price increase does not establish that rivals will join it.
\endgroup

\FloatBarrier\endgroup

\begin{singlespace}
\bibliographystyle{plainnat}
\bibliography{reference}
\end{singlespace}
\end{document}